\documentclass[preprint,showpacs,preprintnumbers,amsmath,amssymb,floatfix]{revtex4}
\usepackage{graphicx}
\usepackage{dcolumn}
\usepackage{bm}
\usepackage{longtable}
\begin{document}

\title{First-principles investigation of the assumptions underlying Model-Hamiltonian
approaches to ferromagnetism of 3$d$ impurities in III-V
semiconductors}
\author{Priya Mahadevan and Alex Zunger \\
National Renewable Energy Laboratory, Golden 80401}
\date{\today}

\begin{abstract}
We use first-principle calculations for transition metal impurities V, Cr, Mn, Fe, Co and Ni in GaAs as well as 
Cr and Mn in GaN, GaP and GaSb to identify the basic features of the electronic structure of 
these systems. The microscopic details of the hole state such as the symmetry and the orbital character, 
as well as the nature of the coupling between the hole and the transition metal 
impurity are determined.
This could help in the construction of model Hamiltonians to obtain a 
description of various properties beyond what current first-principle methods are capable 
of. We find that the introduction of a transition metal impurity in III-V semiconductor introduces a 
pair of levels with $t_2$ symmetry - one localized primarily on the 
transition metal atom referred to as Crystal-Field Resonance (CFR) and the other localized primarily on
the neighboring anions referred 
to as the Dangling Bond Hybrid (DBH). In addition, a set of nonbonding states with $e$ 
symmetry, localized on the transition metal atom are also introduced. Each of the levels is also 
spin-split. Considering Mn 
in the host crystal series GaN $\rightarrow$ GaP $\rightarrow$ GaAs $\rightarrow$ GaSb, 
we find that while in GaN the hole resides in the $t^{CFR}$ level deep in the band gap, in GaAs and GaSb
it resides in the $t^{DBH}$ level located just above the valence-band maximum.
Thus, a DBH-CFR level anticrossing exists along this host-crystal series. A
similar anticrossing occurs for a fixed host crystal ({\it e.g.} GaAs) and 
changing the 3$d$ impurity along the 3$d$ series: V in GaAs represents a DBH-below-CFR limit, whereas Mn corresponds to the 
DBH-above-CFR case. Consequently, the identity of the hole carrying orbital changes. 
The symmetry ($e$ vs. $t_2$), the character (DBH vs. CFR) as well as the occupancy of the
gap level, determine the magnetic ground state favored by the transition metal impurity. 
LDA+U calculations are used to model the effect of lowering the energy of the Mn $d^5$ state 
by varying U. We find that this makes the DBH state more host-like, and at the same time,
diminishes ferromagnetism.
While the spin-splitting of the host valence band in the presence of the impurity has been used
to estimate the exchange coupling between the hole and the transition metal impurity, we show 
how using this would result in a gross underestimation of the coupling.
\end{abstract}


\maketitle
\newpage

\section{Introduction: electronic structure as a tool for constraining model Hamiltonians}
The prospect of manipulating the electron spin to store and transport
information in semiconductor devices has led to renewed interest in
the physics of transition metal (TM) impurities in semiconductors - an area which was
intensively studied in the eighties \cite{alexrev,expt_rev,multiplets_expt}. Current 
interest \cite{expt,dms,nahill,mark,yujun,min,chelikowsky_gan,chelikowsky_gaas,mirbt,HKY,cdgep2,model_dietl1,model_macdonald,model_bhatt,model_amit,model_dietl2}
in achievement of ferromagnetism (FM) at ambient temperatures 
has led to the investigation of the mechanism that stabilizes FM in
transition-metal doped semiconductors. 
One useful approach to obtain an understanding of the electronic properties of
these systems is the {\it first principles electronic structure approach}
where one focuses on the explicit electronic and spin wavefunctions of the system. Variational 
minimization of the total energy, determines within the underlying approximations of the spin density functional
theory some of the basic features of the states involved, such as the extent of localization,
the magnitude of the spin interactions as well as the 
identity of disorder and compensating defects (antisites; interstitials). However, the approach does have the
drawback of underestimating the extent of electron correlations in addition to being a zero-temperature approach.
While comparison with experiment ({\it e.g.} ferromagnetic temperature vs. alloy composition \cite{HKY}) can 
be used to assess the extent to which electron correlations are underestimated, the first-principle 
results have to be generally mapped onto a model Hamiltonian to calculate finite temperature properties.

While model Hamiltonians \cite{model_dietl1,model_macdonald,model_bhatt,model_amit,model_dietl2} have been widely used to
describe the properties of these systems, 
the underlying assumptions in choosing a particular form for the Hamiltonian are rarely justified in their own right.
Generally \cite{model_dietl1,model_macdonald,model_bhatt,model_amit,model_dietl2}, one
renormalizes away the electronic degree of freedom and retains only the spin degree of freedom for the
transition metal impurity. 
One then assumes a local interaction between the transition metal impurity and the free carrier (usually RKKY-like), and
then solves for various physical properties of these systems. 

In the present work we use first-principles 
calculations to examine whether the assumptions made in model Hamiltonian treatments
are consistent with an {\it ab-initio} description of the
electronic structure of these systems.
Our detailed results are then cast in the language of a simple electronic structure model, which could be used 
in an informed construction of model Hamiltonians.

We start by identifying the main physical quantities that come into play in determining the electronic structure
of these systems.
When a trivalent cation site such as Ga of a III-V semiconductor is replaced by 
divalent Mn, an acceptor level (denoted as E(0/-)) is generally created in the band gap. If the Fermi level $\epsilon_F$ lies below
this E(0/-) level, then Mn is charge neutral, {\it i.e.}, its formal 
oxidation +3 equals that of the Ga atom being replaced. In this case there is a hole in the Mn-related orbital.
If, on the other hand, the Fermi level is above E(0/-), then the Mn-related orbital
captures an electron from the Fermi sea ({\it i.e.}
creating a hole there), becoming negatively charged ({\it i.e.} oxidation state Mn$^{2+}$). 
In this case the hole resides in the Fermi sea. 
The Mn-induced hole for $\epsilon_F <$ E(0/-) features prominantly in 
contemporary theories of ferromagnetism 

The model Hamiltonians involve three entities - the host crystal states, 
the transition metal atom, and the impurity-induced hole state. 
There are approximations made at various levels which involve decoupling various
degrees of freedom. At the first level, one decouples the orbital degrees of freedom associated with the
transition metal atom,
describing it with a localized spin-only part. The spin is interacting with a
hole system through a local exchange interaction. At the
next level of approximation, one reduces the problem to that of the transition metal
spin interacting with the hole spin, assuming that the host crystal is unperturbed.
The main assumptions made in such approaches, which we wish to examine, are:

(i) {\it The hole resides in a bulk-like, hydrogenic, delocalized state.} 
This picture is based on the assumption that the perturbing potential $V_{Mn}(r)$-$V_{Ga}(r)$ 
generated by the impurity is dominated by a long-ranged Coulomb part, as a result of 
which only a small percentage of the charge resides in the Wigner-Seitz cell and the rest
is distributed over a large portion of the host crystal. 
In this "host-like hole" picture, one reduces the problem to a  quasi-hydrogenic form
in which the acceptor state is designated via quantum numbers ($s$, $p$ ...) of the host lattice. 
In such cases the wavefunction of the acceptor level is delocalized, and  can essentially
be constructed from the 
host crystal $\Gamma$ states. 
This picture is motivated by the fact that divalent, post transition metal atom elements
such as Zn$^{2+}$ form in III-V semiconductors quasi-hydrogenic acceptor states \cite{baldereschi} 
with small binding energies. 
Similarly, extrinsic $p$-type doping of II-VI dilute magnetic semiconductor CdMnTe  \cite{old_dietl}
also form hydrogenic hole states. However, unlike Zn in GaAs or extrinsic $p$-type CdMnTe, the Mn atom introduced into III-V's 
has chemically active $d$ orbitals \cite{alexrev}, 
so it is not obvious that the acceptor state it forms in GaN, GaP or GaAs would qualify as a delocalized host-like hydrogenic state. Indeed, 
the microscopic features  determining the
localization of the hole wavefunction, such as the $d$-character of the acceptor level
must be considered. Such interactions could change the symmetry ($t_2$ vs. $e$) of the hole state, hence its coupling to the host. The 
pertinent quantum designation of the hole state is impurity-like ($t_2$, $e$) not host-like effective-mass \cite{model_macdonald}.
 
(ii) {\it The host valence band maximum (VBM)  
levels are unperturbed by the transition metal impurity.}
In this view, the host band structure represented in the model Hamiltonian could be described by  a $k$.$p$ model, valid
for the pure host crystal and is decoupled from the part of the Hamiltonian involving the host + hole system.
However, since one of the symmetry representations of the Mn $d$ orbitals in tetrahedral sites
($t_2$, $e$) is the same as that of the zincblende VBM ($t_2$), such states could couple, hence become mutually 
perturbed.

(iii) {\it The spin of the hole couples to the spin of the transition metal impurity via an interatomic
local exchange interaction $J_{pd}$.} As only the spin degree of freedom of the transition metal atom is 
considered, while the orbital degree of freedom is ignored, 
the free carriers feel the effective magnetic field produced by the transition metal impurity spin.
This is modelled as a {\it local} exchange interaction, $J_{pd}$, between the
transition metal impurity and the spin of the free carrier. Hence, the magnitude of $J_{pd}$
determines the energy scale of ferromagnetic ordering. 
Areas visited by the free carrier are rendered ferromagnetic. However certain materials \cite{ferro_insulating} are
found to show activated behavior in their transport implying no free charge carriers, yet they exhibit
ferromagnetism. The current model which requires delocalized carriers cannot explain 
ferromagnetism in such systems.

In what follows, we use first principles calculations to examine the validity of assumptions (i)-(iii) for 
3$d$ impurities V, Cr, Mn, Fe, Co and Ni in GaAs, as well as for Mn and Cr in GaN, GaP and GaSb.
We then construct a qualitative model that explains our numerical 
results. We find that 

(i) The Mn-induced hole could have significant 3$d$  character. 
The assumption of a "delocalized hydrogenic hole" is not supported by first-principle 
calculations. The depth of the acceptor level (reflecting its localization) and the coupling of 
the 3$d$ impurity orbitals to the hole change markedly with the host crystal in the series 
GaN $\rightarrow$ GaP $\rightarrow$ GaAs $\rightarrow$ GaSb.
The hole generated by introducing Mn in GaN is found to have significant 3$d$ character,
while in GaSb the hole is found to have primarily host character. Further, the symmetry
of the hole depends on the combination of the host crystal with the impurity atom. For example
while in GaAs:Co the hole has dominantly $t_2$ character, the corresponding isoelectronic 
impurity ZnSe:Fe  has a hole with $e$ symmetry. 

(ii) The presence of the transition metal impurity perturbs the valence band of the host
semiconductor. The extent of the perturbation depends on the relative position of the 
impurity generated levels  (referenced to the valence band maximum of the host) 
which have the same symmetry as the valence band maximum.
In GaAs:V which has levels with $e$ symmetry in the bandgap, the perturbation is small, while
for GaAs:Mn the perturbation is large. 

(iii) The interaction between the spin of the TM atom and the spin of the host-like hole 
has a predominantly non-local part. This is evidenced by the strong stabilization of the
ferromagnetic state for Mn and Cr pairs in GaAs at $\sim$ 8 $\AA$ separation. 
This interaction induces a spin-polarization of the host-like states.
The direction of the spin-polarization depends on the relative energy position 
of the cation vacancy generated (host-like) states with respect to the impurity states. Furthermore, the
band-theoretic description of Cr in GaP shows a partially occupied 
mid-gap band, and the wavefunctions associated with this mid-gap state are localized.
Yet, even in the absence of free carriers, our total energy calculations predict a ferromagnetic ground state
to be strongly stabilized, while no long range magnetic order is expected.

\section{Earlier Electronic structure Calculations}

There have been considerable earlier first-principles \cite{nahill,mark,yujun,min,chelikowsky_gan,chelikowsky_gaas,mirbt,HKY} 
work on the electronic structure of these systems.
One of the most well-studied systems is Mn in GaAs, which is found to be half-metallic \cite{nahill,mark,yujun,chelikowsky_gaas,mirbt}. 
The GaAs cell with one Mn
atom in it has a net magnetic moment of 4 $\mu_B$ \cite{nahill,yujun,chelikowsky_gaas,mirbt},
with part of the moment residing on the As neighbors of the Mn
atom. The hole state is found to be strongly hybridized with the transition metal state, and has been 
referred to as a hybridized band of holes \cite{chelikowsky_gaas}. The Mn atom and its four nearest neighbors are found to account for
most of the density of states at the valence band edge \cite{chelikowsky_gaas}. Since only the first shell of As 
atoms surrounding Mn are affected
by the spin polarization of the Mn atom, the interactions are believed to be short-ranged \cite{chelikowsky_gaas}. The first 
principle results have been interpreted by Ref.~\cite{nahill} as suggestive of a $d^5$/$d^6$ electron configuration on Mn \cite{nahill}. 
LDA+U calculations \cite{min} have been used to obtain a description of the electronic structure consistent with
photoemission. 

First-principle calculations have been used earlier to elucidate the magnetism in these systems. 
Mahadevan and Zunger \cite{cdgep2} developed a simple model of interaction of the cation-vacancy generated states with
the transition metal states to understand how ferromagnetism results when Mn is doped into the chalcopyrite 
semiconductor CdGeP$_2$. 
Sato and Katayama-Yoshida \cite{HKY} have calculated the energy difference between ferromagnetic and the random alloy
to determine which impurity could give rise to ferromagnetism. They found that at low concentrations V, Cr, and Mn
doping in III-V stabilized the ferromagnetic state, while, Fe, Co and Ni doping stabilized the
magnetically disordered state. Mirbt, Sanyal and Mohn 
\cite{mirbt} showed that the interaction of the transition metal impurity with the As dangling bond states 
could result in a spin polarization of the hybridized dangling bond states. The partial occupancy of these
spin-polarized levels results in ferromagnetism. Sanvito, Ordejon and Hill \cite{nahill}
found an decrease of the spin-splitting of the valence band maximum of the GaAs host with impurity 
concentration. This is in contrast with what is expected from the mean-field Kondo Hamiltonian traditionally used
to describe these systems \cite{kondoiivi}. They attribute the deviation to a breakdown of the
mean-field approximation, while they say that the Kondo Hamiltonian is good enough to 
provide a description of the ferromagnetism. Schilfgaarde and Mryasov \cite{mark} have used the total energies obtained from
first-principle calculations for different materials to extract exchange interaction strengths. They find a
decrease of the exchange interactions with concentration which prompts them to suggest that the picture 
\cite{model_dietl1,model_macdonald,model_bhatt,model_amit,model_dietl2}
of carrier-mediated ferromagnetism is not valid for these systems. 

We build on the current understanding of the electronic and magnetic properties that exists in the literature.
However, we focus our calculations specifically on the examination of the features (i)-(iii) assumed as "input" to most
model Hamiltonian theories.

\section{Method of Calculations}
We have carried out first-principle electronic structure calculations using 
density functional theory, within the momentum space
total energy pseudopotential method \cite{ihmzunger}, using ultra-soft 
pseudopotentials \cite{usp} as implemented in the VASP \cite{vasp} code. 
The Ga pseudopotentials that we used for GaAs and GaP did not include the Ga 3$d$ states in the valence. While 
this is usually a good approximation for GaAs and GaP, it has been found that for GaN not retaining 
Ga 3$d$ states in the valence leads to erroneous results for some physical properties such as 
optimized lattice constant, cohesive energy \cite{fiorentini}. We therefore used ultra-soft 
pseudopotentials which included Ga $d$ states in the basis for GaN.
We studied transition metal impurities - V, Cr, Mn, Fe, Co and Ni in 64 atom supercells of zincblende GaSb, GaAs, GaP and GaN.
In Table I we compare the calculated lattice constants of pure III-V using PW91 GGA exchange functional \cite{pw91} 
with the experimental values \cite{expt_lattice}. We have fixed
the equilibrium lattice constant of the supercells at the 
calculated values of the pure host  given in Table I. The basis sets had a cutoff energy for plane waves equal to 
13.3~Ry for GaSb, GaAs and GaP and 29.4~Ry for GaN. We used a Monkhorst Pack grid of 4x4x4 k-points 
which includes $\Gamma$. The cell-internal positions of the atoms were allowed to relax
to minimize the forces. The equilibrium transition metal-to-As bond lengths in GaAs 
were 2.47, 2.47, 2.48, 2.44, 2.36 and 2.34 $\AA$ for V, Cr, Mn, Fe, Co and Ni respectively.

The $d$ partial density of states as well as the local moment 
at the transition metal were calculated within a sphere of radius
of 1.2 $\AA$ about the atoms and have been broadened with a gaussian of 0.2~eV full width at half maximum.
The total energy differences between the ferromagnetic and the antiferromagnetic 
spin arrangements were computed for TM pairs at first and fourth neighbor separations for 
a parallel and an anti-parallel arrangement of their spins to determine whether a specific 
transition metal impurity resulted in a ferromagnetic state or not.

{\it LDA vs. GGA:} 
In order to compare LDA \cite{capz} and GGA \cite{pw91} 
exchange functionals, we consider the case of
Co impurity in GaAs, where earlier LDA work \cite{mirbt} suggests a nonmagnetic ground state. 
Using the experimental lattice constant of 5.65 $\AA$ for GaAs, 
we find that the GGA calculations lead to a magnetic ground state with a 
moment of 2 $\mu_B$. The energy of  this state is strongly stabilized by $\sim$ 150~meV compared to the
nonmagnetic state. Using LDA exchange functional we find that while the nonmagnetic state is stabilized for a 2x2x2 
Monkhorst Pack grid as used in the earlier work \cite{mirbt}, the magnetic state with the moment of 2 $\mu_B$ is
stabilized by $\sim$ 40 meV for a 4x4x4 Monkhorst Pack k-points grid. 
These observations are consistent with the fact that GGA calculations have a greater ability 
to stabilize a magnetic ground state than LDA calculation. For other impurities - Cr and Mn in GaAs, the LDA and GGA
results are found to give the same ground state. We use the GGA exchange functional throughout this work. 

The introduction of various transition metal impurities lead to defect levels in the band gap of the semiconducting 
host. We compute the formation energies of the transition metal impurities in various charge states $q$.
The formation energy for a defect comprising of atoms $\alpha$ in the charge
state $q$ was computed using the expression \cite{cuinse2}
\begin{eqnarray}
{\Delta}H_f^{\alpha,q}(\epsilon_f,\mu)& =& E(\alpha) - E(0) +
\sum_{\alpha} n_{\alpha} \mu_{\alpha}^{a} + q (E_{v} +
\epsilon_F),
\end{eqnarray}
where  $E(\alpha)$ and $E(0)$ are, respectively the total energies
of a supercell with and without the defect $\alpha$. 
$n_{\alpha}$ denotes the number of atoms of defect $\alpha$
transferred in or out of the reservoir, while $\mu_{\alpha}^a$ denotes their chemical potentials.

{\it Total energies:} The total energies of the
charged supercells were computed by compensating any additional charge on the
impurity atom by a uniform jellium background and 
have been corrected for interactions between charges in neighboring cells 
using the Makov and Payne correction \cite{payne}.
We use the static dielectric constant values - 15.69 for GaSb, 12.4 for GaAs, 
11.11 for GaP and 10.4 for GaN \cite{landolt}.
The quadrupole moment of the isolated defects was calculated as the difference between the moments of the
supercell with the charged defect and that with the neutral defect.

{\it Transition energies:} The defect transition energy $\epsilon(q,q')$ is the value of the Fermi
energy $\epsilon_F$ at which ${\Delta}H^{\alpha,q}(\epsilon_f)$=${\Delta}H^{\alpha,q'}(\epsilon_f)$. The
zero of the Fermi energy is chosen as the valence band maximum
$E_{v}$ of the pure host.

{\it Chemical potential limits:} As the reservoir supplying
the atoms could be elemental solids, or compounds formed from the
elements, we express $\mu_{\alpha}^a$ as the sum of the energy of
the element in its most stable structure $\mu_{\alpha}^s$, and an
additional energy $\mu_{\alpha}$ {\it i.e} $ \mu_{\alpha}^a$ =
$\mu_{\alpha}^s$ + $\mu_{\alpha}$. The stable structures we considered for the elements were 
nonmagnetic body-centred cubic (bcc) for V,
antiferromagnetic bcc for Cr, the antiferromagnetic face-centred cubic (fcc) for Mn, ferromagnetic bcc for Fe,
ferromagnetic hexagonal for Co, ferromagnetic fcc for Ni and the nonmagnetic base-centred orthorhombic 
structure for Ga.

The required ranges of
$\mu_{\alpha}$ are determined by $\mu_{Ga}$ $\le$ 0; $\mu_{TM}$
$\le$ 0; $\mu_{Sb,As,P,N}$ $\le$ 0 (no precipitation of solid elements)
and by the formation energies of the semiconducting host and competing binary phases
formed between the elements of the semiconductor and the transition metal impurity.
This could be the most stable NiAs phase of MnAs in the case of GaAs:Mn and the MnP
phase of CrAs for GaAs:Cr. 

The energies E($\alpha$), E(0), and $\mu_{\alpha}$ entering Eq.(1) are calculated
within the density functional formalism discussed earlier.
No correction for the band gap underestimation was made.
Changing the k-point mesh
from 2x2x2 to 4x4x4 changed the formation energies by $\sim$ 20~meV. 
We used a plane wave cutoff of 13.3 Ry for the calculations involving Mn in GaAs. Increasing the cutoff to 29.4 Ry,
changed the formation energies by $\sim$ 10 meV.

\section{Results of Density Functional Calculations}

We now divide the main features of the first principle calculations into
three main entities introduced in Section I:

A. The nature of the impurity-induced level in the gap

B. The impurity-induced valence-band resonances

C. The perturbed host VBM

Then, in section V, we will provide a simple model that explains all of our numerical results qualitatively.

\subsection{The Nature of the impurity-induced level in the gap}

Figure 1 shows the transition-metal local DOS for V, Cr, Mn, Fe, Co and Ni in GaAs, 
projected into irreducible representations $t_2$ and $e$ and spin 
directions + and -. The VBM is at the zero of the 
energy.  The GGA band gap of pure GaAs is found to be 0.3~eV; all
the impurities V-Ni introduce levels into this band gap. We first 
discuss the nature of these gap levels, and then the circumstances how and when 
a hole is present in them.
\begin{figure}
\includegraphics[width=5.5in,angle=0]{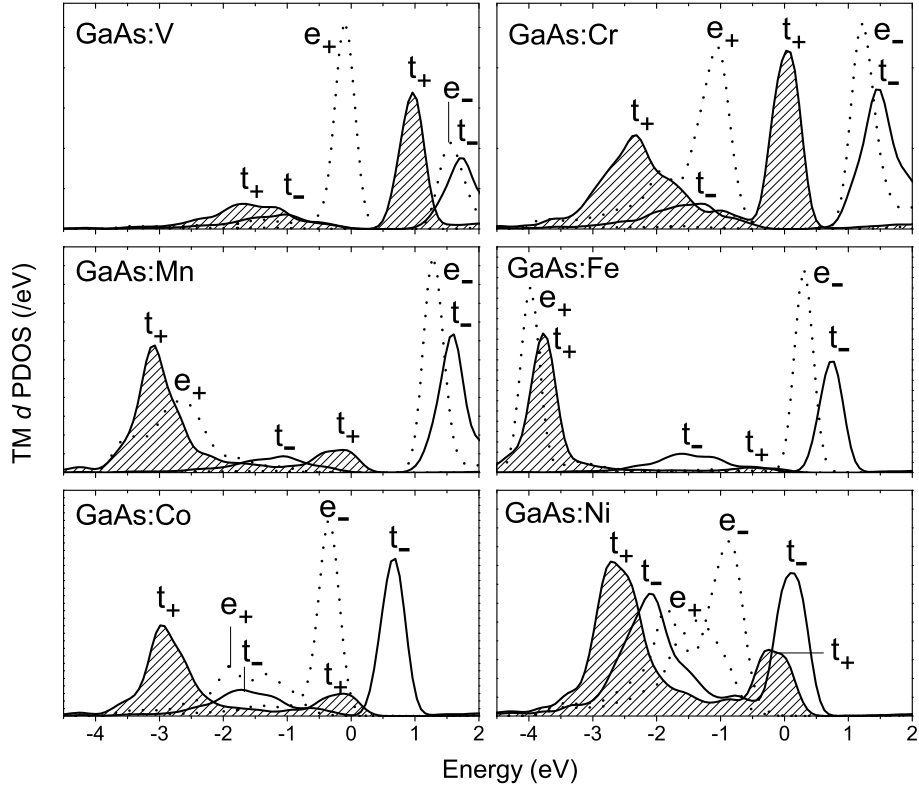}
\caption{ The TM projected density of states for substitutional V-Ni impurities in GaAs evaluated in a sphere of radius 
1.2 $\AA$ for spin-up $t_+$ (shaded region),
spin down $t_-$ (solid black line) and $e$ (dashed line) symmetries. The zero of energy represents the valence band
maximum of the host. The number of k-points used is 64.}
\end{figure}

From Fig.~1 we see that the sequence of levels occupied for Cr in GaAs
are $t_+$, $e_+$, $t_-$ and $t_+$ in the order of increasing energy. For a free atom one would expect levels 
of one spin channel to be filled up before levels of the other spin channel; the deviation that one observes here
reflects solid state effects. The two sets of $t_+$ and $t_-$ levels that we find  for each 
impurity are suggestive of bonding/antibonding combinations arising from hybridization. 
We therefore determined the atoms on which each of the $t_2$ states are localized
by computing atom-projected DOS.
Bonding states with a large wavefunction amplitude on the TM site are referred to 
as "Crystal Field Resonances" (CFR) \cite{alexrev},
whereas antibonding $t_2$ states with low contribution on the TM which are localized instead on the four nearest As atoms
are referred to as the 'Dangling Bond Hybrid' (DBH). The full explanation of the
genesis of these states will be provided in Sec. V. We see that,

1. {\it Symmetry of gap levels and lowest unoccupied levels}: Substituting Cr, Mn and 
Co in GaAs introduce levels with up-spin character and $t_2$ symmetry in the band gap.
These levels are partially occupied by 1, 2 and 2 electrons for neutral Cr, Mn and Co, respectively:
Cr$^0$ ($t_+^1$), Mn$^0$ ($t_+^2$) and Co$^0$ ($t_+^2$). The levels introduced by V$^0$ ($e_+^2$)
and Fe$^0$ ($t_+^3e_+^2$) are fully occupied. The first unoccupied levels have $t_+$ 
and $e_-$ symmetry for V and Fe respectively. 

2. {\it d character of gap levels}: 
The transition metal projected partial density of states for different transition metal impurities
in GaAs given in Fig.~1 indicates that the gap level/first unoccupied level is strongly 
$d$-like for the early transition metal impurities V and Cr, while for the heavier 3$d$ elements
{\it e.g.} Mn these levels have less $d$ character. An increased $d$ character of the gap level would 
imply increased spatial localization of the wavefunction in the vicinity of the impurity. 

\begin{figure}
\includegraphics[width=5.5in,angle=0]{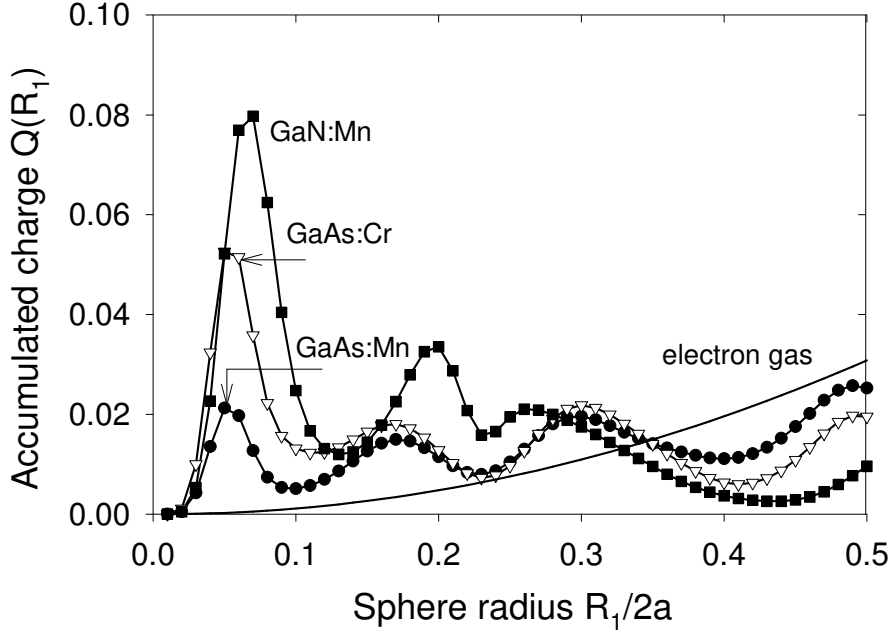}
\caption{ The accumulated charge Q within spheres of radius $R_1$ and $R_1$+$\Delta R_1$ about the TM 
impurity for Mn in GaN (filled squares), Cr in GaAs (open inverted triangles) and Mn in GaAs (filled circles)
compared with the result for an electron gas (solid line). $a$ is the lattice constant of the host 
supercell.}
\end{figure}
3. {\it Degree of localization of gap levels}: We quantify the degree of localization
by plotting in Fig. 2 
the charge Q(R$_1$)=$\int_{R_1}^{R_1+\Delta} \psi^2 r^2 dr$ enclosed between concentric spheres with radius $R_1$ and
$R_1+\Delta$ centered about the 
impurity atom. The integrated charge between the spheres is plotted as a function of R$_1$. 
For comparison, we show also the result expected for a 
homogeneous charge distribution (electron gas), where  the charge density at any point in the cell 
is given by reciprocal volume $\frac{1}{V}$. We see that 
Q(R$_1$) for the TM impurities has little similarity to the results for an
electron gas. Changing the impurity from Mn to Cr in GaAs, we see an increase in the  charge density 
localized in the vicinity of the impurity. We find that till a radius which includes
second neighbors of the TM atom, the integrated charge for Cr is higher than for Mn. Further,  
we find that the enclosed charge in the vicinity of the impurity atom is higher in GaN:Mn than in GaAs:Mn and the decay of the wavefunction is 
faster.

\begin{figure}
\includegraphics[width=5.5in,angle=0]{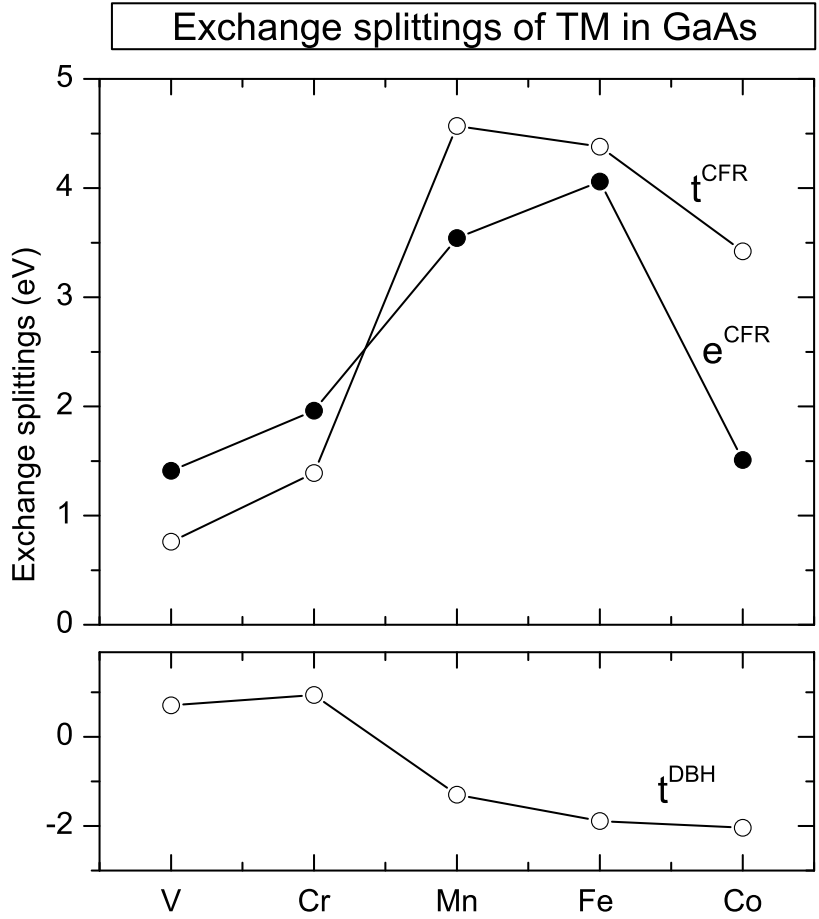}
\caption{ The Exchange-splitting at $\Gamma$ point for 
$t^{CFR}$, $e^{CFR}$ (upper panel) and $t^{DBH}$ (lower panel)  states for V-Co
impurities in GaAs.}
\end{figure}
4. {\it The negative exchange splitting of the gap levels}: Having established the identity of the gap levels, 
we now investigate their spin splittings. 
In Fig. 3 we have plotted the spin splittings of the CFR and DBH levels at
$\Gamma$ point for the impurities
V-Co in GaAs obtained from an analysis of their 
eigenvalues/eigenfunctions. For V and Cr the spin splitting of the DBH levels is
positive {\it i.e.}
$t_+^{DBH}$ states are at lower energies compared to $t_-^{DBH}$. However, for
Mn, Fe and Co the splitting is negative with the
$t_-^{DBH}$ states at lower energies compared to $t_+^{DBH}$.
A similar negative exchange splitting was observed earlier for the Te states in 
MnTe \cite{mnte}, the Ce states in CeFe$_2$ \cite{cef2}, and the Mo states in Sr$_2$FeMoO$_6$ \cite{sfmo}.
The explanation was that the $p$ states of Te in CdTe, the $d$ states of Ce in CeFe$_2$ and the $d$ states
of Mo in Sr$_2$FeMoO$_6$ are sandwitched in between the 3$d$ states of the transition metal atom. The $p$-$d$
hybridization results in an exchange splitting of these states opposite in direction to that on the transition
metal atom. Indeed we see from Fig. 1 that for the cases where the DBH states
are bracketed by the spin split CFR states, the spin splitting is negative.

5. {\it Enhanced exchange splitting for $t^{CFR}$ states}: From Fig. 3, we see
that the exchange splitting of the $t^{CFR}$ states is larger than that of  
the $e^{CFR}$ states for Mn, Fe and Co impurities. 

Having discussed the existence of impurity-induced levels in the band gap
of the host semiconductor, we next discuss the location of these levels.

6. {\it Acceptor transitions for gap levels}: 
Single particle LDA or GGA levels do not have any rigorous meaning. We thus calculate
{\it transition} energies, $\epsilon$($q$,$q'$) which correspond to the value of the Fermi energy $\epsilon_F$
at which the defect changes from a charge state $q$ to $q'$.
Table II provides the calculated and measured  \cite{multiplets_expt} acceptor/donor transition energies for various transition 
metal impurities in GaAs.
The calculated acceptor levels for Mn and Cr in GaSb, GaAs, GaP and GaN are plotted in Fig. 4 where the host band
edges are aligned according to their calculated unstrained valence band offsets \cite{boffset}.
We see that as the electronegativity of the host crystal increases in the sequence GaSb $\rightarrow$ GaAs $\rightarrow$ GaP $\rightarrow$ GaN, 
its bulk ionization energy (= position of VBM with respect to vacuum) increases. 
The acceptor level is thus farther away from the 
VBM of GaN than it is from the VBM of GaAs. Thus, GaN:Mn and GaP:Cr have more localized hole states whereas GaSb:Mn 
has more delocalized holes.
This behavior, whereby the acceptor energy level does not follow the host valence band energy (as in the
case of hydrogenic impurities) characterizes localized states \cite{fazzio}.
\begin{figure}
\includegraphics[width=5.5in,angle=0]{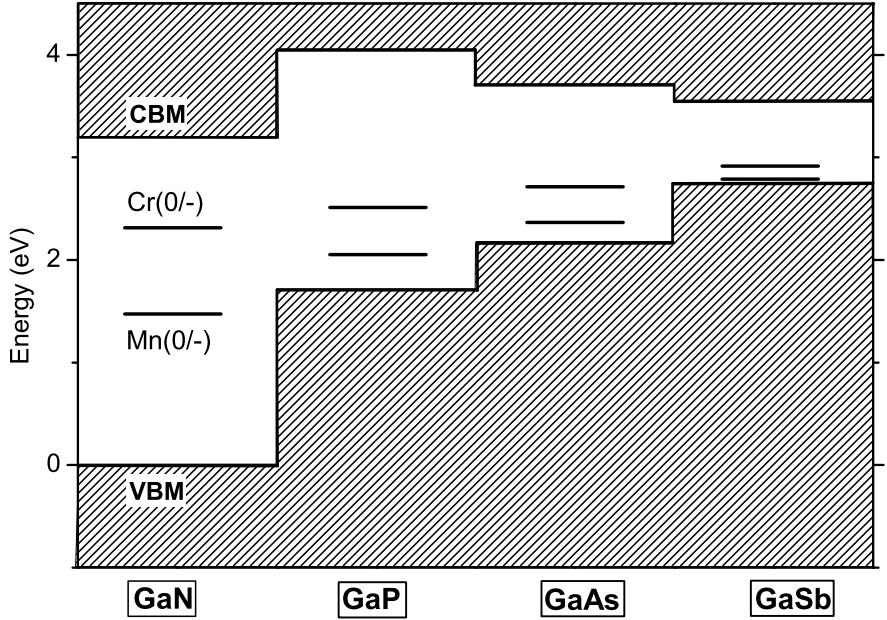}
\caption{ The (0/-) acceptor transition energies for Cr and Mn impurities in GaN, GaP, GaAs and GaSb.
The band edges of the host semiconductors are aligned according to the LDA-calculated unstrained band
offsets \cite{boffset}, and the gaps are the experimental values.}
\end{figure}

7. {\it Multiplet states and violation of isovalency rule}: We use the level occupancies (Fig. 1) 
as well as the net magnetic moments that
we obtain for different transition metal impurities in GaAs (Table III) 
to obtain the multiplet configuration describing the ground state. 
These are given in Table III. We also provide the multiplet configuration observed from experiment \cite{multiplets_expt} 
and find that there is
agreement in all cases. It is interesting to compare the ground state multiplets of two isoelectronic 
cases - ZnSe:Fe$^{2+}$ and GaAs:Co$^{3+}$ in their neutral charge states. 
In both cases we expect an electron configuration of $d^6$.
Normally, one would expect to find equal multiplets for isoelectronic cases \cite{alexrev} (isovalency rule).
This expectation is based on the fact that we are looking at a low transition metal impurity concentration regime where
basic crystal field theory ideas are expected to be sufficient to explain the observed ordering of energy levels.
However, we find in GaAs:Co$^{3+}$ the configuration $^3$T$_2$, i.e the hole is in the $t^{DBH}$ level 
($t_{CFR+}^3 ~e_{CFR+}^{2} ~e_{CFR-}^{2} ~t_{DBH-}^3{\bf ~t_{DBH+}^2}$), 
whereas in ZnSe:Fe$^{2+}$ we find $^5$E, i.e the hole is in the $e_{CFR}$ level 
($t_{CFR+}^3 ~e_{CFR+}^{2} ~t_{DBH-}^3 ~t_{DBH+}^3 {\bf ~e_{CFR-}^{1}}$). 
The reason for the difference is that the stronger $p$-$d$ hybridization for GaAs:Co
pushes the $t_+^{DBH}$ levels to higher energies, so that the $e^{CFR}_-$ levels are occupied first and the
hole resides in the $t_+^{DBH}$ level. 
We thus conclude that the isovalency rule is not applicable, and one cannot
assume that the hole is in a "generic" $d$ state. 

8. {\it FM vs. AFM ground state and their relation to the symmetry of the gap levels}:
Having summarized the nature of the level induced in the gap by the introduction of the transition metal 
impurity, we now analyze when a ferromagnetic state is favored. In Table III we provide the energy
difference between the ferromagnetic and antiferromagnetic energies for two TM atoms at nearest $\Delta$E$_{NN}$ and 
fourth neighbor $\Delta$E$_{4NN}$ fcc positions in a 64 atom supercell of GaAs. We find that (a) when the level 
in the gap is fully occupied as in V$^0$ and Fe$^0$, the favored ground state is antiferromagnetic. (b) When the level 
in the gap is partially occupied and has $t_2$ symmetry as in Cr$^0$, Mn$^0$, the ferromagnetic state is lower in energy.
This is also the case for electron doped V$^-$ in GaAs which is strongly ferromagnetic.
Although Co$^0$ ($^3$T$_2$) also has a hole in the $t_2$ level, the system is at the brink 
of a ferromagnetic-to-nonmagnetic transition.
(c) When the level in the gap has $e$ symmetry, as in the case of electron doped Fe$^-$, the stability of the
FM state is weaker. 
Evidently the {\it symmetry} of the hole carrying state strongly determines the magnetic order.
{\it We conclude that FM is stabilized strongly only when the hole resides in the level with $t_2$
symmetry.} (Note , viz. Sec. V, that in $T_d$ symmetry $t_2$ states are strongly bonded to their neighboring
atoms, whereas the lobes of the $e$ orbitals point in between the nearest-neighbor atoms.)

\subsection{The impurity-induced valence-band resonances}

Figure 1 shows that in addition to the gap levels, the introduction of a transition metal atom gives rise to 
resonance levels that lie deep within the valence band of the host semiconductor. 
In most model Hamiltonian theories 
\cite{model_dietl1,model_macdonald,model_bhatt,model_amit,model_dietl2}, one usually ignores the
orbital degree of freedom of the transition metal impurity,  and
the presence of the impurity is included only as a localized spin of value 5/2.
In our calculations we find that the degree of localization of such deep
resonances (thus, the possibility of depicting them as local point-like spin)
varies sharply with the position of the impurity in the periodic table.
For heavier TM such as Fe and Mn, the deeper resonance
level has significant TM $d$ character (being Crystal-Field Resonances), while for the early TM impurities in GaAs, 
one finds that the deeper $t_2$ 
levels have significantly less TM character (being Dangling Bond Hybrids). This is discussed next.

1. {\it Anticrossing of the two $t_2$ levels in different host materials:} 
Level anticrossing is evident when keeping the impurity atom fixed, and, changing the
host semiconductor. Considering the example of Mn, we find that by changing the host from GaSb to GaN,
the DBH and the CFR exhibit anticrossing. 
This is not the only difference: 
We find that the exchange splitting of the DBH levels is in the same direction as the CFR levels 
(positive) in GaN:Mn, in contrast to GaAs:Mn. Further,
in GaN:Mn the $t^{CFR}_+$ levels lie above the $e^{CFR}_+$ levels, unlike the case in GaAs:Mn.
The reason is evident from Fig. 4 which shows that the VBM of GaN
is much deeper than the VBM of GaAs. Since the free Mn$^{2+}$ ion has its $d$ orbitals {\it above} the GaN VBM, 
but {\it below} the VBM of GaSB or GaAs, an
anticrossing occurs along the GaN $\rightarrow$ GaP $\rightarrow$ GaAs $\rightarrow$ GaSb series. This is illustrated in Fig. 5
which shows that in GaN:Mn for the up spin channel, the upper $t_2$ is more localized than the lower $t_2$, whereas in GaAs:Mn the 
localization sequence is reversed.
This clarifies a confusion that existed in the literature \cite{dietl_rev} regarding the question of whether the gap level is 
localized or not. Our result shows that the answer depends on the host. These results also clarify the
nature of the acceptor transition for Mn in different materials. GaN:Mn can be viewed as a $d^4$-like case
since its configuration is $e_{CFR+}^2 t_{CFR+}^2$ (hole in $t^{CFR}_{+}$),
and the (0/-) acceptor transition is from a Mn configuration $d^4$ to $d^5$. On the other hand 
for Mn in GaP, GaAs and GaSb we have the configuration ($d^5$ + hole), {\it i.e.}
$e_{CFR+}^2 ~t_{CFR+}^3 ~t_{DBH-}^3 ~t_{DBH+}^2$ and the 
acceptor transition is from a Mn configuration of ($d^5$+hole) to $d^5$.
\begin{figure}
\includegraphics[width=5.5in,angle=0]{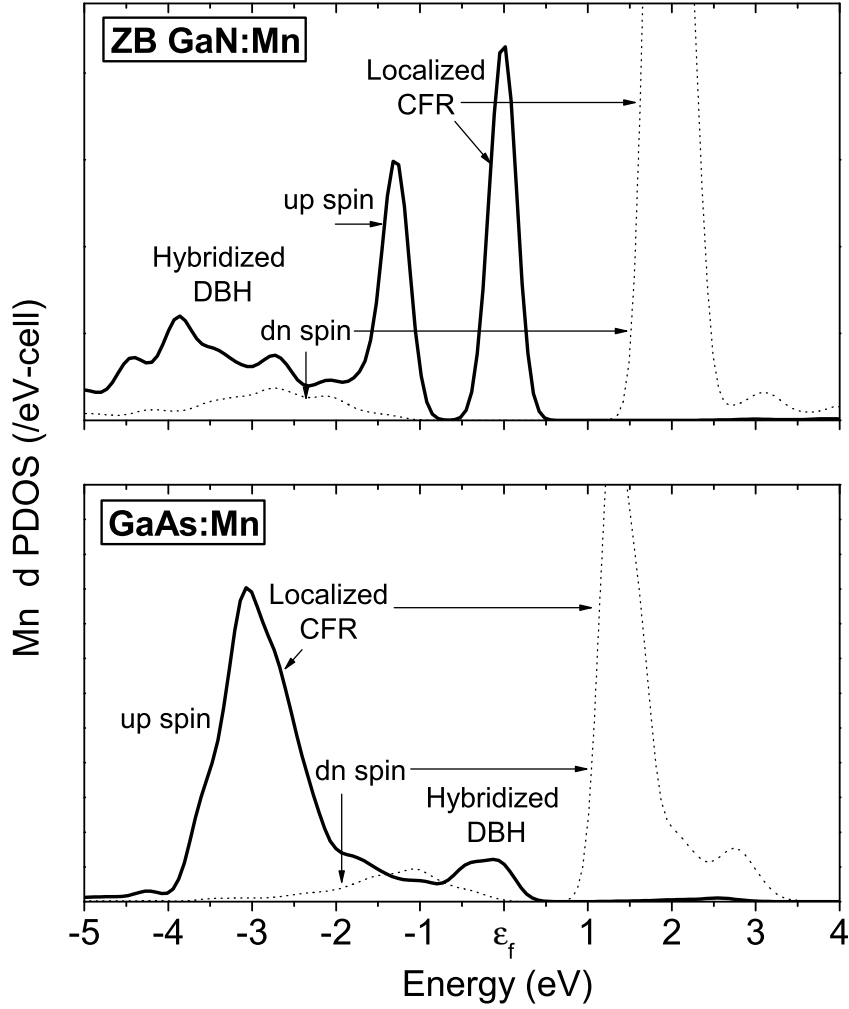}
\caption{ The up and down spin Mn $d$ projected partial density of states evaluated within a sphere of radius 1.2 $\AA$ for a Mn impurity in 
GaN (upper panel) and GaAs (lower panel). The number of k-points used is 64. }
\end{figure}

2. {\it Occupancy of the valence band resonances and comparison with photoemission}: Table III gives the calculated 
occupancy of the crystal field resonances of the 3$d$ impurities in GaAs (in square brackets). These levels are found to have
a configuration of "$d^5$" for Cr, Mn and Fe. 
Experimentally the position of these levels can be detected by valence band photoemission \cite{fujimori}. By 
suitably tuning the photon energy so that the photo-ionization cross-section is maximum for the TM-related states, an electron 
can be ionized from these deep CFR levels. Kobayashi {\it et al.} \cite{fujimori} used resonant valence band photoemission 
and showed that the the CFR levels for Mn in GaAs are located at $E_v$ - 4~eV. A direct comparison of the position of these
levels with the single-particle density of states calculated for GaAs:Mn places these 
energies at $E_v$ - 2-3~eV. The LDA error in the position of these
states is because of the self-interaction correction (SIC) 
that places these energies too high \cite{sicref}. As pointed out earlier \cite{zungeriivi} for the
3$d$ states in II-VI's, the experimental result should be compared with the total energy difference between the configurations 
$d^4$ and $d^5$ and not with the bare single particle eigenvalues. Alternatively,
the LDA error can be empirically 
corrected by using the simplified LDA+U version of SIC. In Fig. 6 we plot the Mn $t_+^{CFR}$  
partial density of states as a function of U for GaAs:Mn. As $U$ increases, the position of the 
Mn $t^{CFR}$ level is pushed deeper into the GaAs valence band. Agreement with XPS 
for the $t_+^{CFR}$ being at $E_v$-4~eV occurs for U $\sim$ 2~eV. At the
same time, the Mn character
of the DBH state at $\epsilon_F$ (not shown) decreases with increasing $U$. 
As the DBH hole becomes more delocalized, the E$_{FM}$-E$_{AFM}$ stabilization energy (insert to Fig. 6) is reduced; 
a "host-like hole" obtained for unphysically large U leads to nearly vanishing FM stabilization energy. Clearly, the 
picture of "host like hole" is invalid for GaAs:Mn, since for the $U$ that leads to agreement with XPS the DBH hole is still 
localized to some extent, whereas for very large $U$, when the hole is delocalized, there is no ferromagnetism.
\begin{figure}
\includegraphics[width=5.5in,angle=0]{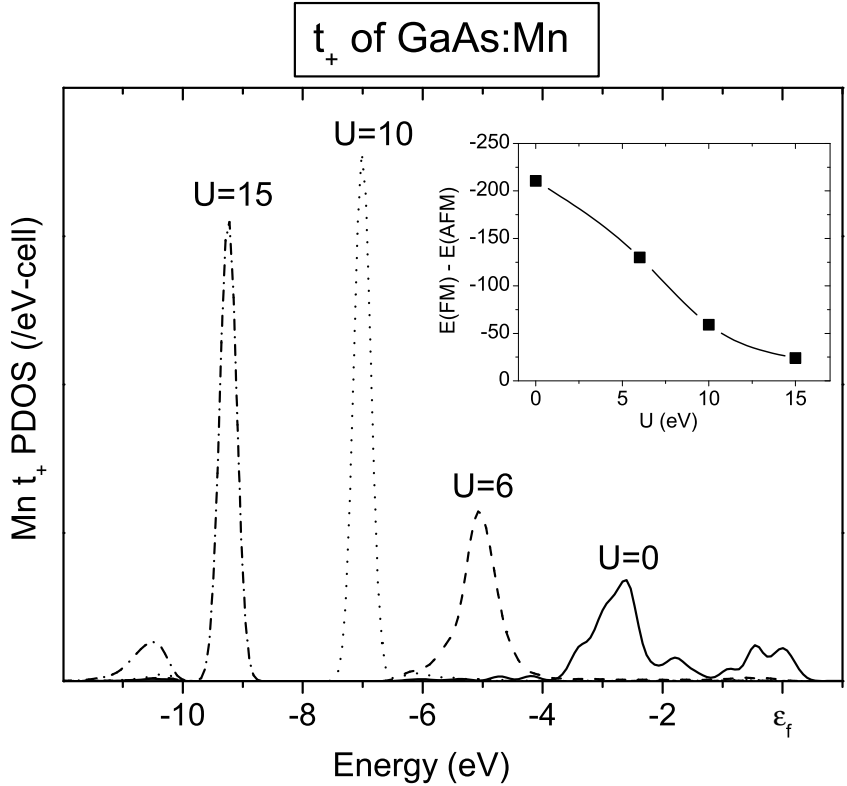}
\caption{ Mn $t_+$ projected partial density of states evaluated within a sphere of radius 1.2 $\AA$ 
for on-site Coulomb interaction strengths 
U=0, 6 and 10 and 15 for Mn in GaAs. The number of k-points used is 4x4x4.
The inset shows the variation in E(FM)-E(AFM) for two Mn atoms at nearest neighbor positions.}
\end{figure}

\subsection{The perturbed host VBM}

\begin{figure}
\includegraphics[width=4.5in,angle=270]{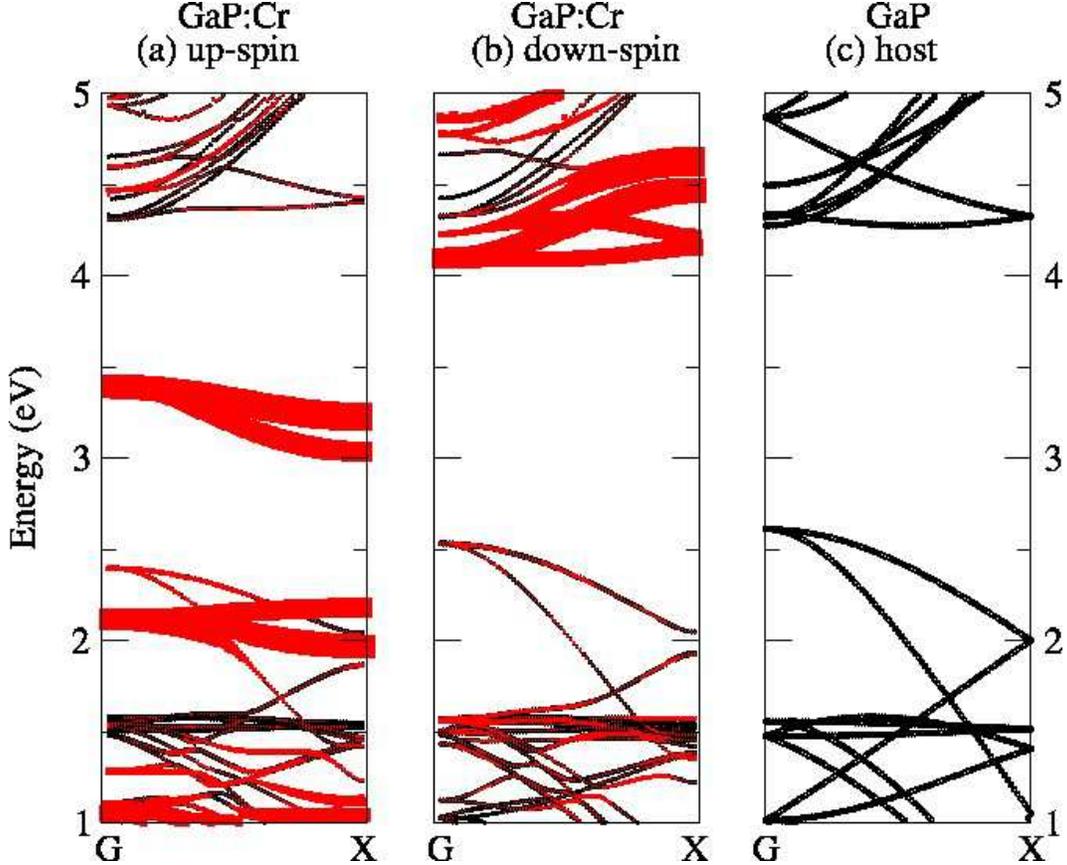}
\caption{ The band dispersions in (a) up and (b) down spin channel for GaP:Cr compared with (c) the host 
supercell. The thickness of the lines represents the Cr weight in the bands.}
\end{figure}
Having studied the impurity-induced levels in the gap and deep in the host valence band, we next examine the perturbation of
the host states, especially the host valence band maximum by the presence of the impurity atom. Figure 7 shows the
up and down spin band dispersions for 3 $\%$ Cr doped GaP 
supercell (panels (a) and (b)). The band dispersion of the GaP host without the
impurity has been provided in panel (c) for comparison. The
thickness of the lines depicting these bands has been made proportional to the Cr $d$ character of the states. 
We see that Cr introduces a new band within the band gap of GaP. In a band-theoretic picture, this system
is metallic, with the Fermi energy within the impurity band. Interestingly, (1)
the host band dispersions 
are significantly altered by the presence of the impurity. In particular the VBM is found to have
significant TM $d$ character for the 3$\%$ Cr concentration represented by the supercell.
(2) A Cr-induced spin-splitting of the valence band maximum is observed. Effects (1) and (2) suggest that the 
host VBM is sufficiently perturbed by the transition metal.

\begin{figure}
\includegraphics[width=5.5in,angle=0]{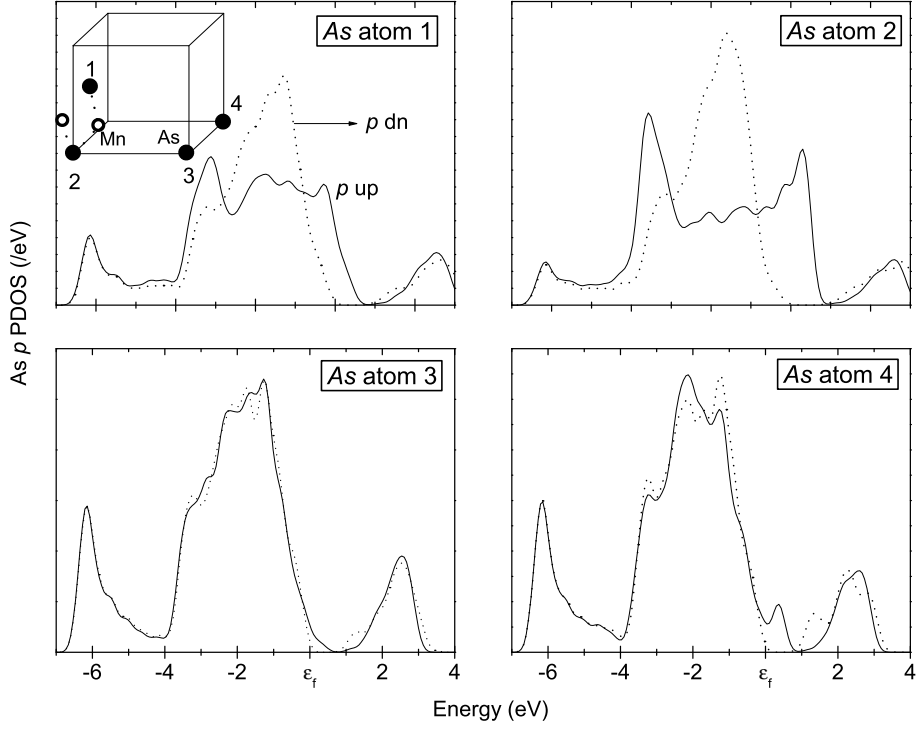}
\caption{ Up (solid line) and down (dashed line) spin projected partial density of states for As atoms labelled
1-4 in Mn substituted GaAs 
evaluated within spheres of radius 1.2 $\AA$ using 64 k-points. 
The positions of the As atoms (filled circles) with respect to the Mn atoms is shown in the inset.}
\end{figure}
Another way of detecting perturbations in the host bands 
is to examine the host projected DOS of the system containing the impurity.
In Fig. 8 we have plotted the As $p$ partial density of states projected
onto different As atoms labelled 1-4 for a GaAs supercell containing
2 Mn atoms.
The As atom labelled 1 has one Mn nearest neighbor, while the As atom labelled 2
has two Mn nearest neighbors. The As atoms
show a strong polarization which increases with the number of Mn neighbors.
The As atoms labelled 3 and 4 which are far away from the Mn atoms show a
reduced polarization.

To pictorially see the perturbation in the VBM states, 
we compare in Fig. 9 the wavefunction squared along two chains in the (110) plane for the 
valence band maximum of the pure GaAs host (panel (a)) as well as the 
VBM ({\it i.e.} state below DBH) of the system with the Mn
impurity in the up (panel (b)) and down (panel (c)) spin channels. 
The upper chain in panels (b) and (c) contains the perturbing Mn impurity.
The perturbation of the VBM in the presence of the impurity can be assessed 
by comparing the perturbed charge density for each spin channel 
with the unperturbed charge density of the host lattice. We find that the perturbations are significant 
in the chain containing the Mn atom, while in adjoining chains, the perturbation is limited in extent. 
Further the perturbations are stronger in the up spin channel than in the down spin channel.
\begin{figure}
\includegraphics[width=3.5in,angle=0]{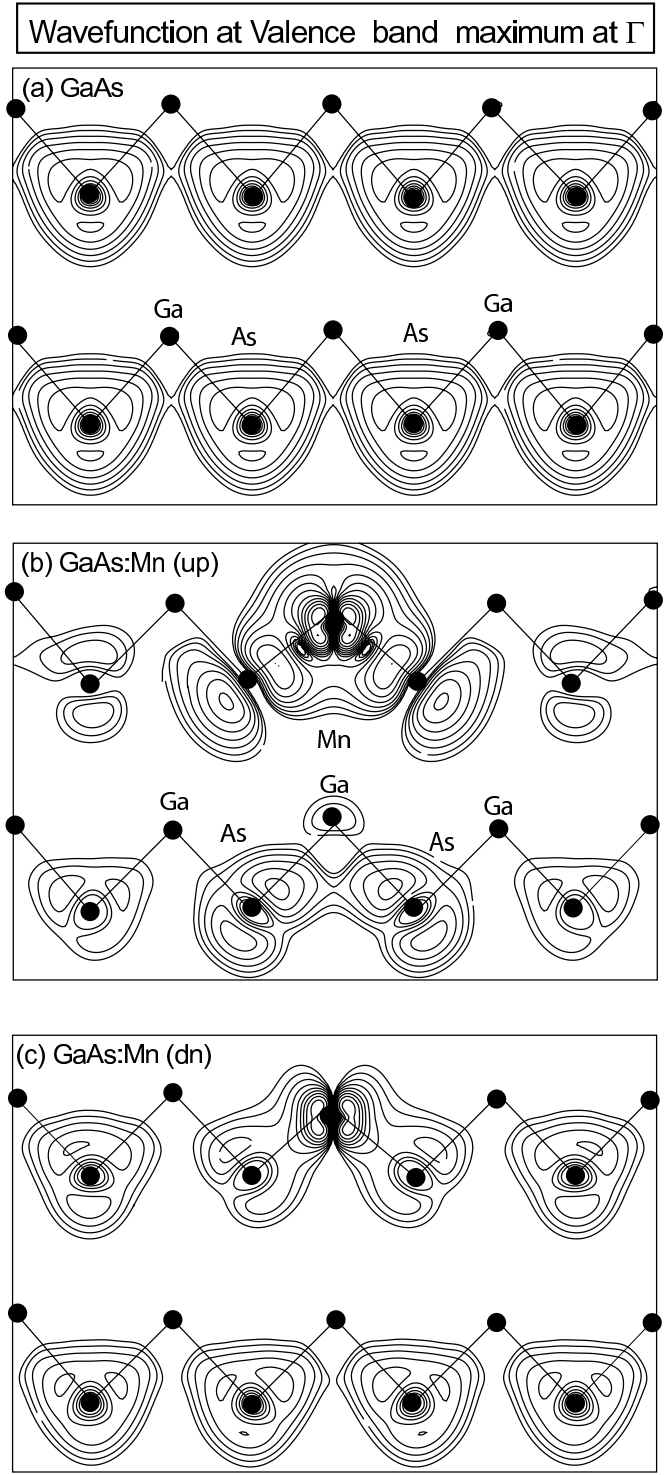}
\caption{ The wavefunction squared at the valence band maximum for (a) the pure host GaAs 
(b) for GaAs:Mn in the up spin channel and the (c) down spin channels. The lowest contour 
corresponds to 0.0015 e/$\AA^3$ and each contour is 1.6 times larger.
}
\end{figure}

To evaluate the 3$d$-induced spin-splitting in the VBM, 
we reference the up and down spin VBM eigenvalues 
of the impure system to the corresponding VBM of the pure host semiconductor.
This is done by aligning the average potentials on Ga atoms far away from the impurity for the two systems.
The presence of the impurity band with $t_2$ symmetry above the VBM
for Mn and Cr impurities complicates the identification of the valence band maximum. 
We associate the highest occupied triply degenerate eigenvalues at $\Gamma$ point with the 
impurity band, and the next deeper set as $E_v^+$.
The shift with respect to the pure host
$\Delta E_{v}^+$=$E_{v}^+$(GaAs:Mn)-$E_{v}$ (GaAs) and
$\Delta E_{v}^-$=$E_{v}^-$(GaAs:Mn)-$E_{v}$ (GaAs)
is given in Table IV for the impurities V, Cr and Mn in different host 
semiconductors. We find that the perturbation of the host VBM is smaller in the down spin channel compared 
to the up spin channel. This is consistent with what we find from the charge density plotted in Fig. 9.
The spin splitting of the valence band maximum depends strongly on the transition metal impurity and
host semiconductor combination. Considering the case of impurities in GaAs, we find that while the spin splitting
associated with the introduction of V is only 0.06 eV, it increases to 0.39 eV for Mn. Keeping the
impurity fixed (Mn), and varying the semiconductor host (GaAs to GaN), we find the splitting decreases from 0.39 eV to 0.1 eV.
The small valence band splittings in the case of V in GaAs as well as Mn in GaN compared with that for Mn in GaAs is 
because of the larger energy separation between the interacting $t_2$ states in the former cases compared to the latter.

\subsection {Summary of the electronic structure as obtained by density functional}

We are now in a position to examine whether the physical picture of the electronic structure of 3$d$ impurities as 
assumed in model Hamiltonian theories (reviewed in Sec. I) is consistent with first-principles calculations (outlined
in Sec. IV).

(i) {\it The nature of the TM-induced hole state}: 
A 3$d$ impurity in a III-V semiconductor generates two sets of states with $t_2$ symmetry, 
and one set of states with $e$ symmetry in each spin channel. While one set of $t_2$ states are localized 
on the TM atom (CFR), the other are localized on the host anion atoms next to the impurity (DBH). These states CFR and DBH
exhibit anticrossing for a fixed TM as a function of the host anion GaN $\rightarrow$ GaP $\rightarrow$ GaAs $\rightarrow$ GaSb, 
or for a fixed host as a function of the impurity V $\rightarrow$ Mn. 
The localization of the hole state decreases as we move from Mn in GaN to Mn in GaP, and then to Mn in GaSb.
Not all impurities introduce holes. In GaAs, 
V$^0$ and Fe$^0$ have no hole; Cr$^0$, Mn$^0$ and V$^-$
have $t_2$ holes; and Fe$^-$ has an $e$ hole. 
In all cases, however, the hole is non-hydrogenic, manifesting significant
admixture of 3$d$ character and showing deep acceptor levels whose 
energies do not follow the host VBM. 
This implies that the neglect of the short-range part of the impurity 
potential and the consequent expansion of the acceptor wavefunction in terms 
of a single host wavefunction are questionable. The effective mass of the 
hole state is therefore different from that of the host, as observed in recent experiments \cite{optical_prl}. 
The exchange splitting of the CFR states is different for the $t_2$ states from that for the $e$ states. 
While the splitting for the $e$ states is larger than that for the $t_2$ states for V and Cr in GaAs, the order is reversed
for Mn, Fe and Co. This reversal in the order of the spin splitting of the CFR states is accompanied by a
reversal in the sign of the spin splitting of the DBH states. The identity of the hole state - both the symmetry as well as
the character depends on the impurity-host combination. While the hole carrying orbital for Fe in ZnSe has $e$ symmetry, 
the hole is found to be located in an orbital with $t_2$ symmetry for the isovalent doping of Co in GaAs.

(ii) {\it The nature of the host VBM}: The introduction of the transition metal perturbs the valence band of the host 
crystal. We find this perturbation to be large when the state in the gap has $t^{DBH}$ character. 
This is because the effective coupling is larger since the DBH states have strong host character.
We find that the VBM is spin-split in the presence of 3$d$ impurity, and that the VBM in the up-spin channel is
perturbed more strongly than in the spin down channel in the presence of the impurity.

(iii) {\it Ferromagnetism and symmetry}: Impurities with fully occupied DBH-like $t_2$ gap states such as V$^0$, 
Fe$^0$ show antiferromagnetism. Partial occupation of the $t^{DBH}$ as in Cr$^0$, Mn$^0$ or V$^-$
show ferromagnetism. Partially occupied $e$-like level like in Fe$^-$ show weak or no ferromagnetism. 

(iv) {\it Ferromagnetism and hole localization}: 
Using LDA+U as an artificial device to explore the consequences of delocalized host-like-hole states we find
(insert to Fig. 8) that in this limit there is reduced ferromagnetism.

We find that despite the well known GGA-LDA band gap error, as well as the underestimation 
of the location of deep CFR states due to SIC, these first principle calculations
provide us with the correct spin multiplets. LDA+U changes some details (CFR locations), but does not
alter the basic picture emerging from GGA/LDA when the hole is DBH-like as in GaAs:Mn.

\section{ Simple Model of the Electronic Structure of 3$d$ impurities in GaAs}

\subsection{The Model}

Most of the results of the density functional study of the electronic structure of 3$d$ impurities in III-V's (Sec III)
can be captured by a simple model. The electronic structure of substitutional 3$d$ in III-V semiconductors can
be understood as arising from the interaction of the host cation vacancy 
(= anion dangling bonds) with the crystal-field and exchange split orbitals of a 3$d$ ion.

(a) {\it The dangling bonds for a column III cation vacancy $V_{III}$}:
A cation vacancy for a column III element gives rise to a fully-occupied $s$-like $a_1$ level
located deep in the host valence band, 
and a partially-occupied $p$-like $t_2$ level located just above
the host valence band maximum, with wavefunction amplitude localized primarily
on the neighboring atoms \cite{baraff}. 
This is evident from the wavefunction squared
of the Ga vacancy dangling bond state shown in the (110) plane in Fig. 10.
The neutral vacancy $V_{III}^0$ has a deficiency of 3 electrons, {\it i.e} the orbital configuration
is $a_1^2t_2^3(p)$, where $p$ denotes its major orbital character.
Spin-polarization splits this $t_2(p)$ vacancy level into spin-up [$t_+(p)$] and
spin-down [$t_-(p)$] states, but the splitting is small (90~meV at the $\Gamma$ point) on account of the rather
delocalized nature of these pure host dangling bond orbitals. 
\begin{figure}
\includegraphics[width=5.5in,angle=0]{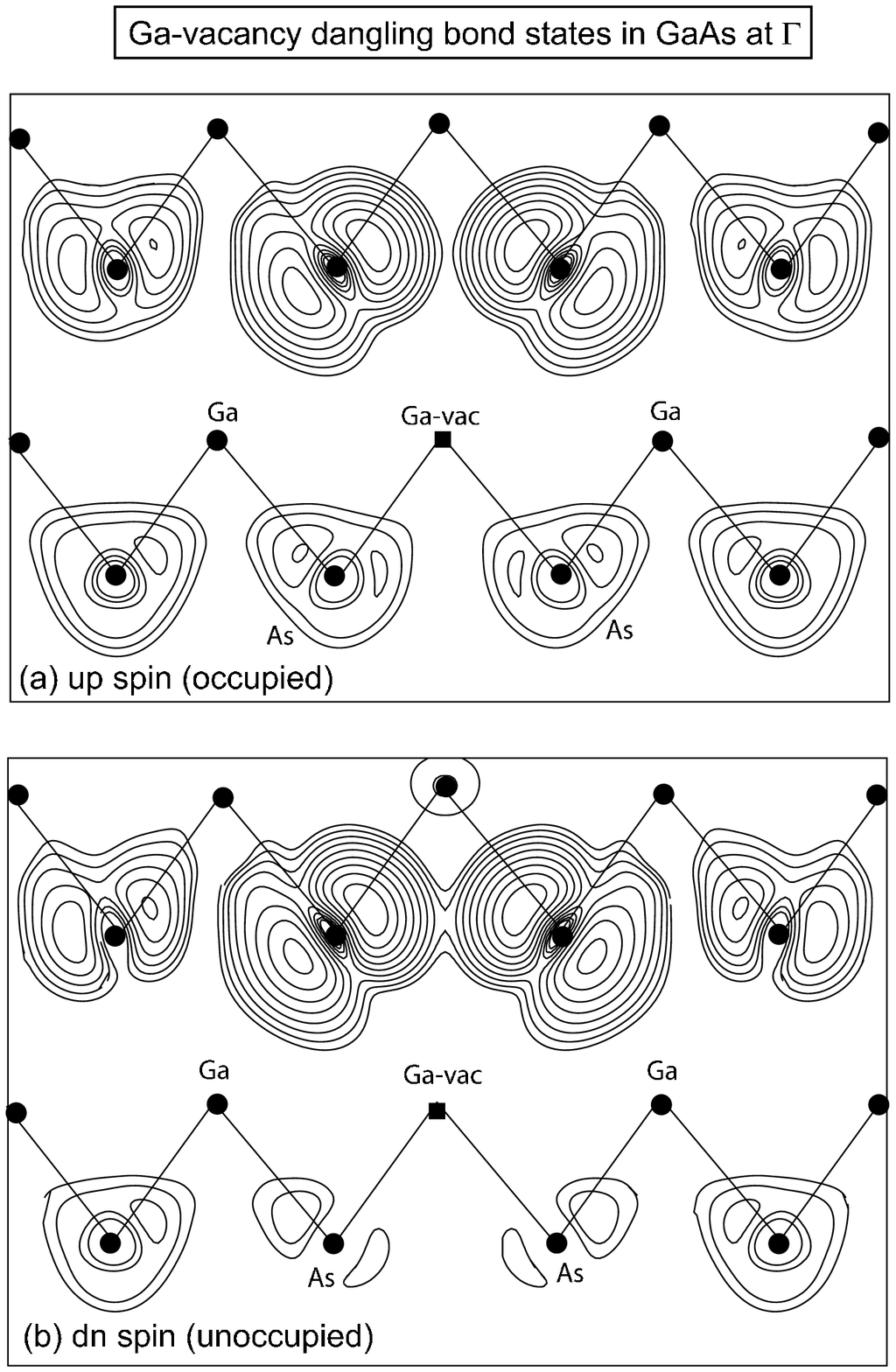}
\caption{ The wavefunction squared for the (a) up and (b) down spin Ga-vacancy generated dangling bond states 
with $t_2$ symmetry in pure GaAs. The lowest contour corresponds to 0.0017 e/$\AA^3$ and each contour is 1.6 times larger. }
\end{figure}

(b) {\it The crystal-field split TM 3$d$ orbitals}:
The tetrahedral crystal-field of the zincblende host
splits the TM $d$ levels into $e(d)$ and $t_2(d)$, with $e$ below
$t_2$ in the point-ion limit \cite{balhausen}; the crystal-field (CF)
splitting of the ion is denoted by $\Delta_{CF}$($t_2$-$e$). Exchange interactions further
split these levels into spin-up~(+) and spin-down~(-), with exchange-splittings
$\Delta_x(e)\equiv$[$e_-$(d)-$e_+$(d)] and $\Delta_x(t)\equiv$[$t_-$(d)-$t_+$(d)]. 

\begin{figure}
\includegraphics[width=5.5in,angle=270]{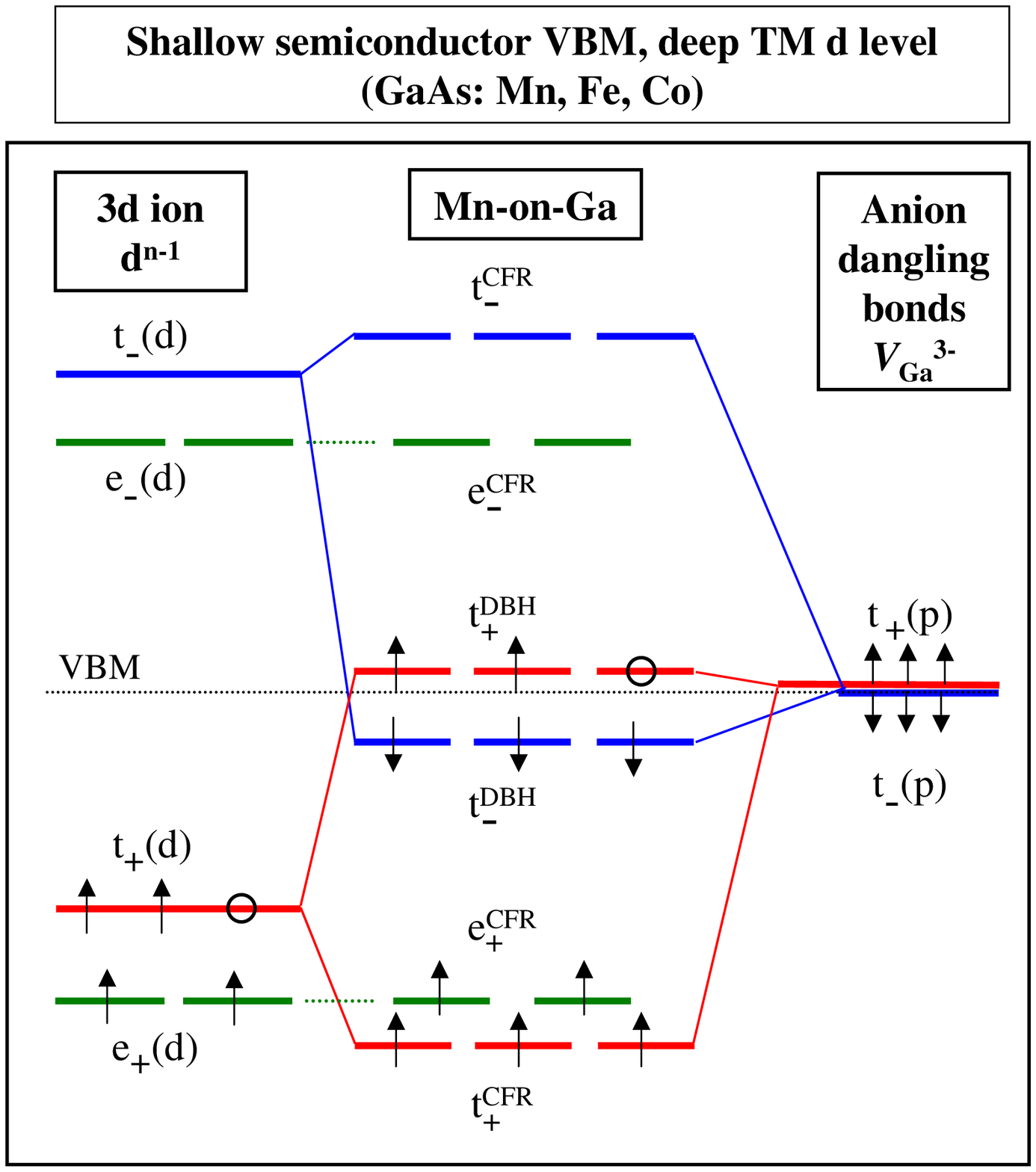}
\caption{ The schematic energy level diagram for the levels (central panel) generated from 
the interaction between the crystal-field and exchange-split 
split levels on the  3$d$ transition metal ion (left panel) with the anion dangling bond levels (right panel), 
when the TM $d$ levels are energetically deeper than the dangling bond levels.}
\end{figure}
\begin{figure}
\includegraphics[width=5.5in,angle=270]{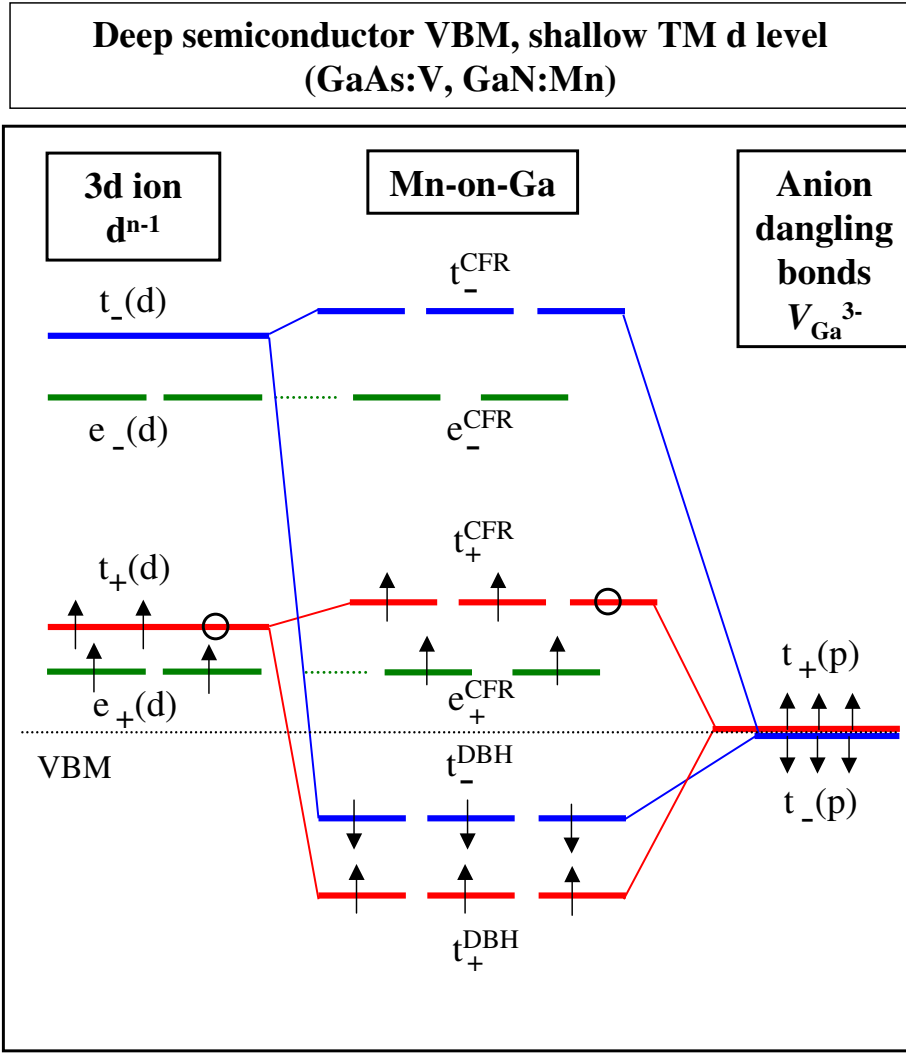}
\caption{ The schematic energy level diagram for the levels (central panel) generated from 
the interaction between the crystal-field and exchange-split 
split levels on the  3$d$ transition metal ion (left panel) with the anion dangling bond levels (right panel), 
when the TM $d$ levels are energetically shallower than the dangling bond levels.}
\end{figure}
The energy levels of a cation-substituted TM in a III-V semiconductor 
can be thought of \cite{alexrev} as the result of hybridization between
the anion dangling bonds generated by a column III cation vacancy $V_{III}$ [(i) above], and the 
crystal-field and exchange split $d$ levels
of a TM ion placed at the vacant site [(ii) above]. 
There are two limiting cases: When the 3$d$ levels are well below the host cation dangling bonds 
(e.g. Mn in GaAs, Fig. 11), or when the 3$d$ levels are well above the host cation dangling bonds
(e.g. V in GaAs, Fig. 12).
The dangling bond states are shown on the right hand side of Fig.~11 and 12, while 
the crystal field and exchange split 
TM $d$ levels are shown on the left hand side of Fig.~11 and 12. The levels generated after hybridization are shown 
in the central panel.
The $t_2(p)$ levels of the anion dangling bond 
hybridize with the $t_2(d)$ levels of the transition metal. In contrast, the $e(d)$ level
of the TM ion remains largely unperturbed since the host does not have
localized $e$  states in this energy range, available for significant 
coupling. 
Considering the examples of Mn in GaAs and GaN, we plot the charge density of the $e$ and the $t_2$
states in the (110) plane in Fig. 13. It is evident that the $e$ states for Mn in GaAs are essentially nonbonding, while in GaN, 
as a result of the reduced Mn-Mn separation, there is weak interaction between the Mn atoms.
The hybridization in the $t_2$-channel creates
bonding, transition-metal localized "Crystal-Field Resonances" (CFR):
$t^{CFR}_+$, $t^{CFR}_-$, as well as the host-anion localized
antibonding "Dangling Bond Hybrids" (DBH): $t^{DBH}_+$, $t^{DBH}_-$, whereas the
$e$-channel creates the non-bonding $e^{CFR}_+$ and $e^{CFR}_-$ states. 
This model explains the existence of two sets of $t_+$ and $t_-$ levels that we found in Fig.~1.
The available electrons for occupation of these levels are N=(n-1)+6 for a $d^n$s$^2$ transition metal
atom (three electrons are used to complete the anion dangling bond state to $t_2^6$($p$), leaving
$d^{n-1}$ at the transition metal ion.)
For GaAs:V (Fig. 12) the ordering of levels after hybridization is 
$$t_+^{DBH} < t_-^{DBH} <  e^{CFR}_+ < t_+^{CFR} < e_-^{CFR}< t_-^{CFR}$$ with increasing energy. 
Hence for V we have N=(n-1)+6=8 electrons occupy the DBH and CFR levels. 
Thus, V has the configuration  
($t_{DBH+}^3 t_{DBH-}^3 e_{CFR+}^{2}$), 
as seen in Fig. ~1 and Table III.
\begin{figure}
\includegraphics[width=5.5in,angle=0]{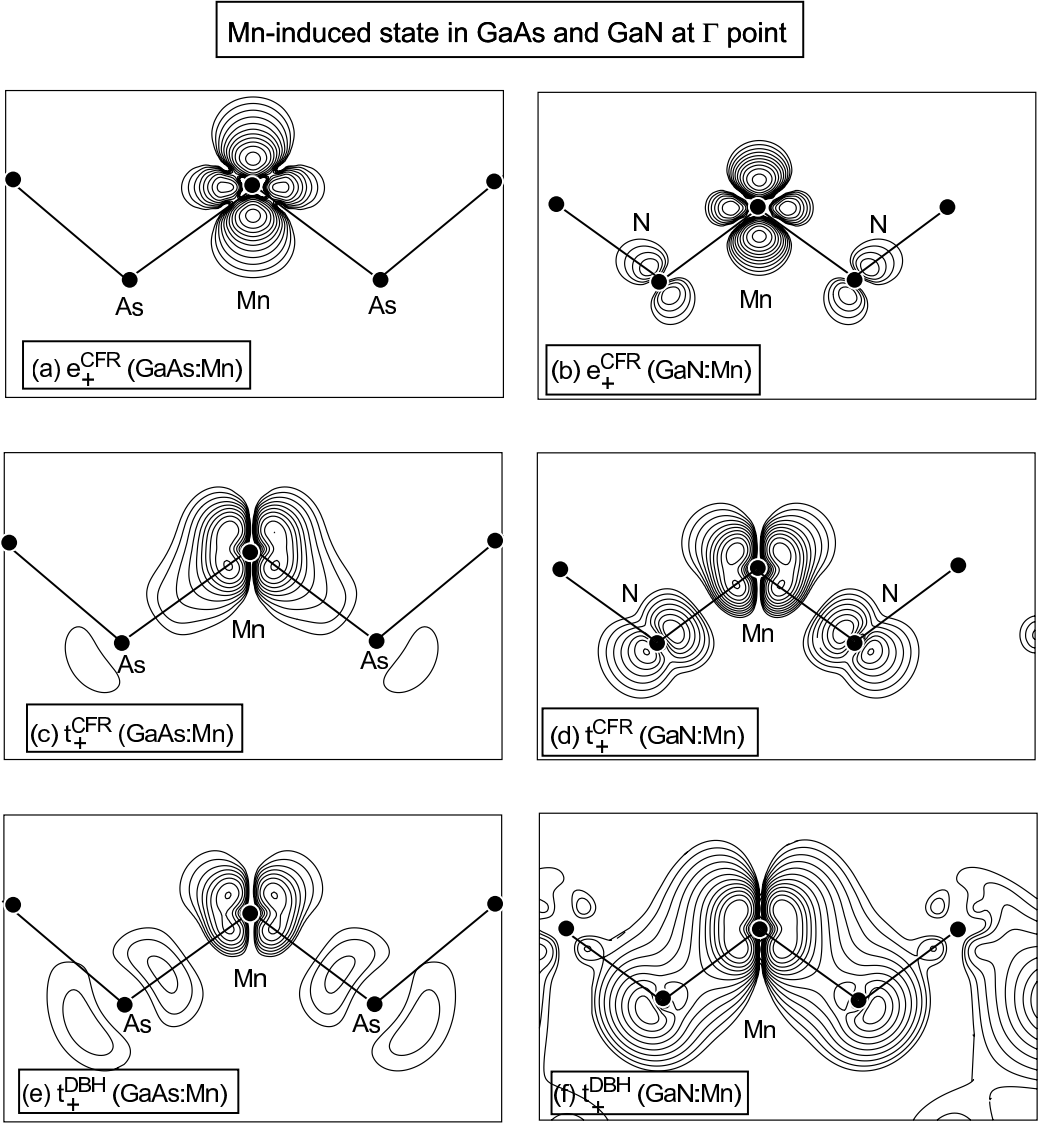}
\caption{ The wavefunction squared of Mn induced (a) $e^{CFR}$ in GaAs:Mn, (b) $e^{CFR}$ in zinc blende GaN:Mn, (c) $t^{CFR}$ in GaAs:Mn, 
(d) $t^{CFR}$ in zinc blende GaN:Mn, (e) $t^{DBH}$ in GaAs:Mn and (f) $t^{DBH}$ in zinc blende GaN:Mn.
The lowest contour corresponds to 0.015 e/$\AA^3$ and each contour is 1.6 times larger.}
\end{figure}

The order of levels for
$$ \mbox{GaAs:Mn~~ is~~} (t_{CFR+}^3 < e_{CFR+}^{2} < t_{DBH-}^3 < {t_{DBH+}^2}) \mbox{,}$$
$$ \mbox{GaAs:Fe~~ is~~} (t_{CFR+}^3 < e_{CFR+}^{2} < t_{DBH-}^3 < {t_{DBH+}^3}) \mbox{,}$$ 
$$ \mbox{GaAs:Co~~ is~~} (t_{CFR+}^3 < e_{CFR+}^{2} < e_{CFR-}^{2} < t_{DBH-}^3 < {t_{DBH+}^2})$$ 
(Fig. ~11). The number of electrons (n-1)+6
is 10, 11 and 12 for Mn, Fe and Co respectively. 
This agrees with Fig.~1 showing that Mn and Fe in GaAs have the ordering of levels shown
in Fig.~11, with fully-filled $t_+^{CFR}$ 
and $e_+^{CFR}$ levels and 2, 1 and 0 holes in the $t_+^{DBH}$ level.
By an analysis of the density of states obtained within our first principle calculations, 
we have determined (Table III) the energy minimizing orbital configurations for the transition metal 
impurities V, Cr, Mn, Fe and Co in GaAs in fully relaxed configurations. The first 
unoccupied orbital for each impurity has been indicated in boldface in Table III.
The simple model of Figs. 11, 12 gives the same result.

\subsection{Qualitative consequences of the simple model}

\begin{figure}
\includegraphics[width=5.5in,angle=270]{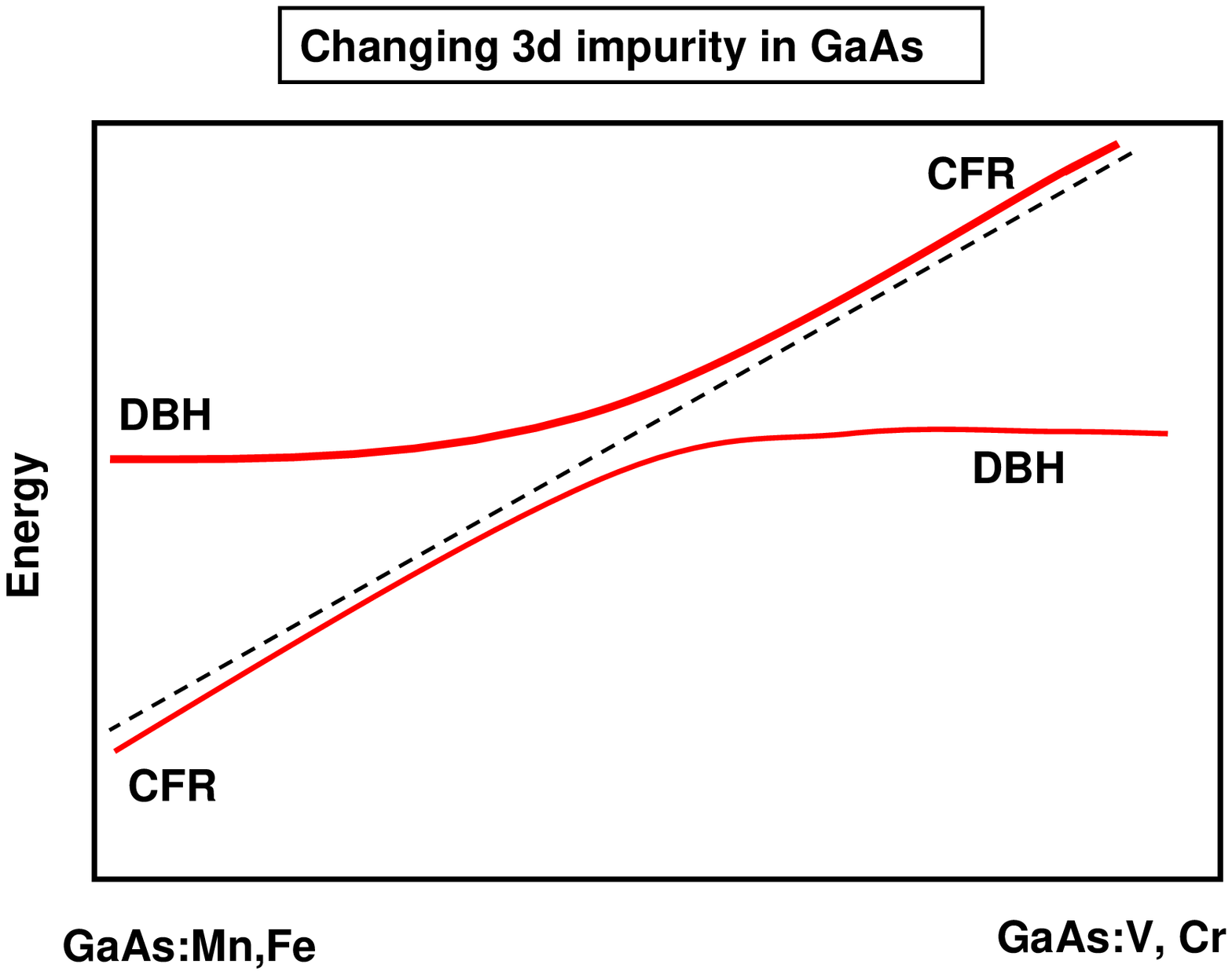}
\caption{ The schematic plot of band anticrossing between the two $t_2$-like levels in GaAs for different 3$d$ impurities.
}
\end{figure}

1. {\it Level anticrossing}: 
The model explains how the hopping interaction between the $t_2$ states on the transition metal 
impurity with the cation-vacancy states generates a pair of $t_2$ states in each spin channel.
The bonding-antibonding character of these states is determined by the relative separation of the
interacting levels as well as their interaction strengths. 
Hence as depicted in Fig. 14, one could by a suitable choice of the TM impurity change the character of the
gap levels. When the orbital energy of the 3$d$ ion lies below the host dangling bond, we have a "CFR-below-DBH"
situation, illustrated in Fig. 11. In this case one has 
CFR states in the valence band of the semiconductor while the gap levels are more delocalized 
with dominant weight in the dangling bonds. This is 
the case for GaAs:Fe, Mn, and Co. Conversely, when the orbital energy of the 3$d$ ion lies {\it above}
the host dangling bond, we have the "CFR-above-DBH" situation, illustrated in Fig. 12. In this case the
gap level is CFR-like.

While Fig. 1 illustrates anticrossing when changing the 3$d$ atom, but keeping the host fixed, 
{\it e.g.} GaAs, Fig. 5 suggests that there is also anticrossing when keeping the 3$d$ atom fixed
{\it e.g.} Mn, but changing the host crystal GaN $\rightarrow$ GaP $\rightarrow$ GaSb. Indeed, 
as the anion dangling bond becomes deeper (GaSb $\rightarrow$ GaAs $\rightarrow$ GaP $\rightarrow$ GaN) 
or the TM $d$ level becomes shallower (Mn $\rightarrow$ V), there will be a DBH-CFR anticrossing,
resulting in the level ordering shown in Fig.~12. The CFR-above-DBH is exemplified
by GaAs:V (see Fig.~1) which has a shallow $d$ level and by GaN:V, Cr and Mn which have a deep
semiconductor VBM \cite{boffset}. 
For GaAs:V, this level anti-crossing results in a reduced ${\Delta_{x}}$($t_2$) exchange 
splitting compared to the cases where this crossing did not occur (Cr and beyond).
For Mn in III-V's we see a reversal between hole-in-DBH case for GaAs:Mn with
$[t^3_+e_+^2t_-^0e_-^0]_{CFR}$ ~$({\bf t_+^2}t_-^3)_{DBH}$ configuration to the case of hole-in-CFR
$[{\bf t_+^2}e_+^2t_-^0e_-^0]_{CFR}$ ~$(t_+^3t_-^3)_{DBH}$ for GaN:Mn. As the CFR is much more 
localized than DBH, this is reflected by  a change in acceptor level depth. 

2. {\it ~Negative exchange splitting in the DBH manifold and low-spin
configurations}: For Co,Fe and Mn in GaAs the special position
of the Ga-vacancy level $t_{\mp}$(p) {\it between} the exchange-split TM ion levels
$t_+$(d) and $t_-$(d) (Fig. 11), results in a hybridization-induced 
exchange splitting of the DBH states (Fig.~3), opposite in 
direction to the splitting on the CFR levels (Fig.~3). This results from the fact 
that $t_-^{DBH}$ is pushed down by $t_-$($d$)-$t_-$($p$) coupling more than 
the $t_+$($d$)-$t_+$($p$) coupling pushes $t_+^{DBH}$ up. Such 
a negative exchange splitting was also observed in bulk MnTe \cite{mnte},
CeF$_2$ \cite{cef2} and Sr$_2$FeMoO$_6$ \cite{sfmo}.
As seen in Table III, the negative exchange splitting leads to the DBH orbital configuration
($t^3_-{\bf t^2_+}$) and ($t_+^3t_-^3$) for Mn and Fe, respectively. 
This corresponds to a  low-spin configurations for Mn with a moment of
4. Co, on the other hand, has a configuration of
[$t^3_+$$e^2_+$$e^2_-$]$_{CFR}$~($t_-^3{\bf t_+^2}$)$_{DBH}$, so $\mu$=2$\mu_B$.
Note that the moment $\mu$=4 for GaAs:Mn is a consequence of 
$\Delta_x < 0 $: the $e^2_+t^3_+$ CFR levels give a spin of 5/2, and the $t_-^{DBH}$-below-$t_+^{DBH}$ 
gives a configuration $t_-^3t_+^2$, so the total spin is 4/2, and $\mu$=4. Had $t_-^{DBH}$ been above
$t_+^{DBH}$, we would have had the configuration 
$t_+^3t_-^2$ with S=3, and $\mu$=6. As a result of the negative exchange splitting of the DBH states, one finds that
the exchange splitting for the more delocalized $t^{CFR}$ levels is larger than that for the more localized $e^{CFR}$
levels.

3. {\it ~Migration of the d-holes into the DBH acceptor states of GaAs:Cr, Mn and Co}:
The $d^{n-1}$ configuration  of the trivalent TM ion corresponds to $d^3$ for Cr and $d^4$
for  Mn in GaAs and thus to 2 and 1 hole respectively, relative 
to $d^{5}$. However, since for these impurities the $t_+^{CFR}$ and $e_+^{CFR}$ levels are deeper in energy
than the $t^{DBH}$ levels (Fig.~11), it is energetically favorable
for the $d$ holes to "float" into the higher $t^{DBH}$ levels. As a result,
we find that GaAs:Cr ($d^3$) has a configuration $[t_+^3 e_+^2]_{CFR}$$(t^3_-t_+^1)_{DBH}$
with two holes in the DBH, while for GaAs:Mn ($d^4$) we find $[t_+^3 e_+^2]_{CFR}$$(t^3_-t_+^2)_{DBH}$
with one hole, and GaAs:Fe ($d^5$)
has a closed-shell configuration of $[t_+^3e_+^2e^0_-]_{CFR}$$(t^3_+t^3_-)_{DBH}$. 
One expects Co ($d^6$) to have 
a configuration $[t_+^3e_+^2e_-^1]_{CFR}$$(t_-^3t_+^3)_{DBH}$, {\it i.e.} a ground state multiplet $^5E$ with a hole in $e^{CFR}$. 
However, we find that the lowest energy configuration
for Co is $[t_+^3e_+^2e_-^2]_{CFR}$$(t_-^3t_+^2)_{DBH}$ 
with a hole in the DBH level {\it i.e.} ground state multiplet $^3T_2$. 
The migration of the holes from the TM-localized deep CFR levels into the dangling-bond manifold for Cr, Mn and Co
in GaAs creates partially occupied $p$-$d$ hybridized states at the Fermi level with significant delocalized As $p$
character. 

4. {\it Perturbation of the valence band maximum of the host}: 
The valence band maximum of the pure host has $t_2$ symmetry. Hence, they can interact 
with the states with the same 
symmetry on the dangling bonds as well as the TM ion. The consequent spin splitting of the VBM must depend on the 
relative separation of the levels involved as well as the coupling strength. 
Our analysis for different TM-host combinations suggests that when the gap level has $t^{DBH}$  character
the spin-splitting of the VBM is large ($\sim$ 0.4~eV for Mn in GaAs). 
Changing the host semiconductor from GaAs to GaN, increases the energy separation between the dangling bond 
levels generated by the Ga vacancy and the valence band levels. Consequently the perturbation of the
valence band levels resulting in the observed spin-splitting is smaller ($\sim$ 0.1~eV for Mn in GaN). 
This VBM spin-splitting of the impure system has been traditionally used to 
estimate the exchange interaction strength $J_{pd}$ between the 
hole and the transition metal atom. We see that the spin splittings of the VBM 
of the impure system is 
grossly underestimates the spin-splitting of the \underline{impurity band} (DBH,CFR) as seen in Figs. 1,3 and 5.

\section{Conclusions}

We have analyzed the basic electronic structure of 3$d$ transition metal impurities in III-V semiconductors. We find that the
introduction of a 3$d$ impurity is accompanied by the introduction of a pair of states with $t_2$ symmetry in addition to 
nonbonding states with $e$ symmetry. Not all 3$d$ impurities introduce holes. The basic symmetry and character of the hole state
depends strongly on the semiconductor-impurity combination. We find that the hole has significant 3$d$ character.
We have constructed a microscopic model which captures the basic aspects of the electronic structure of transition metal 
impurities in semiconductors. The elements of this model are the relative separation of the dangling bond and transition metal
levels, the $p$-$d$ hybridization strength and the crystal-field and exchange splittings of the transition metal levels.
We model a change in these interaction strengths by changing the semiconducting host, keeping 
the transition metal impurity fixed, - Mn. We find that while the hole introduced by Mn has 
significant 3$d$ character in GaN, it is more delocalized in GaAs. 
The symmetry ($e$ vs. $t_2$), the character (DBH vs. CFR) as well as the occupancy of the
gap level, determine the magnetic ground state favored by the transition metal impurity. 
When the hole has dominantly host character, an
exchange splitting is induced on the hole states which is opposite in direction to that on that the transition metal atom.
The nature of the exchange coupling $J_{pd}$ that exists between between the transition metal atom and the hole comes out
automatically from such a microscopic model. The perturbation of the host valence band is not directly related to the coupling strength
$J_{pd}$. When the hole has primarily DBH character, one finds the perturbation of the host valence band is larger.
The basic picture that emerges from our first-principles calculations could be used to replace the more naive
model-Hamiltonian treatments which have assumed a host-like-hole picture, an unperturbed valence band and that the
spin of the hole couples to the spin of the TM via a local exchange interaction.

This work was supported by the U.S. DOE, Office of Science, BES-DMS under contract no. DE-AC36-99-G010337.


\begin{thebibliography}{}

\bibitem{alexrev}
{\it See} A.~Zunger in {\it Solid State Physics}, Edt. F.Seitz, H.~Ehrenreich 
and D.~Turnbull vol. {\bf 39}, 275 (Academic Press, New York, 1986) and
references therein.

\bibitem{expt_rev}
{Semiconductors and Semimetals Vol. 25}, Edt. J.K. Furdyna and J. Kossut, (Academic Press 1988);
C. Delerue, M. Lannoo and G. Allan, Phys. Rev. B {\bf 39}, 1669 (1989).

\bibitem{multiplets_expt}
J.~Schneider in {\it Defects in Semiconductors II, Symposium Proceedings}, Edt. S. Mahajan and J.W. Corbett, 
225 (North-Holland, 1983);
B.~Clerjaud, J. Phys. C {\bf 18}, 3615 (1985).

\bibitem{expt}
H.~Ohno, Science {\bf 281}, 951 (1998);
Y.~Matsumoto, M. Murakami, T. Shono, T. Hasegawa, T. Fukumura, M. Kawasaki, P. Ahmet, T. Chikyow, S. Koshihara
and H. Koinuma, Science {\bf 291}, 854 (2001);
M.E.~Overberg {\it et al.}, Appl. Phys. Lett. {\bf 79}, 3128 (2001);
M.L.~Reed {\it et al.}, Appl. Phys. Lett. {\bf 79}, 3473 (2001);
Y.D. Park, A.T. Hanbicki, S.C. Erwin, C.S. Hellberg, J.M. Sullivan, J.E. Mattson, T.F. Ambrose, A. Wilson, G. Spanos
and B.T. Jonker, {\bf 295} 651 (2002);
H. Saito, V. Zayets, S. Yamagata and K. Ando, Phys. Rev. Lett. {\bf 90}, 167202 (2003).

\bibitem{dms}
V.I.~Litvinov and V.K.~Dugaev, Phys. Rev. Lett. {\bf 86}, 5593 (2001);
J. Fernandez-Rossier and L.J. Sham, Phys. Rev B {\bf 64}, 235323 (2001);
J. Inoue, S. Nonoyama and H. Itoh, Phys. Rev. Lett. {\bf 85}, 4610 (2000);
G. Zarand and B. Janko, Phys. Rev. Lett. {\bf 89}, 047201 (2002);
G. Alvarez, M. Mayr and E. Dagotto, Phys. Rev. Lett. {\bf 89}, 277202 (2002);
Y.J. Zhao, T. Shishidou and A.J. Freeman, Phys. Rev. Lett. {\bf 90}, 047204 (2003). 

\bibitem{nahill}
S. Sanvito, P. Ordejon and N.A. Hill, Phys. Rev. B {\bf 63}, 165206 (2001).

\bibitem{mark}
M. van Schilfgaarde and O.N. Mryasov, Phys. Rev. B {\bf 63}, 233205 (2001).

\bibitem{yujun}
Y.J.~Zhao, W.T. Geng, K.T. Park and A.J. Freeman, Phys. Rev. B {\bf 64}, 035207 (2001).

\bibitem{min}
J.H. Park, S.K. Kwon and B.I. Min, Physica B {\bf 281-282}, 703 (2000).

\bibitem{chelikowsky_gan}
L. Kronik, M. Jain and J.R. Chelikowsky, Phys. Rev. B {\bf 66}, 041203 (2002).

\bibitem{chelikowsky_gaas}
M. Jain, L. Kronik, J.R. Chelikowsky and V.V. Godlevsky, Phys. Rev. B {\bf 64}, 245205 (2001).

\bibitem{mirbt}
S. Mirbt, B. Sanyal and P. Mohn, J. Phys. Condens. Matter {\bf 14}, 3295 (2002).

\bibitem{HKY}
K. Sato and H. Katayama-Yoshida, Semicon. Sci. and Tech. {\bf 17}, 367 (2002).

\bibitem{model_dietl1} 
T.~Dietl, H. Ohno, F. Matsukura, J. Cibert and D. Ferrand, Science {\bf 287}, 1019 (2000).

\bibitem{model_macdonald} 
J. K$\ddot{o}$nig, H. Lin, A.H. MacDonald, Phys. Rev. Lett. {\bf 84}, 5628 (2000).

\bibitem{model_bhatt}
M. Berciu and R.N. Bhatt, Phys. Rev. Lett. {\bf 87}, 107203 (2001).

\bibitem{model_amit}
A. Chattopadhyay, S.~Das Sarma and A.J.~Millis, Phys. Rev. Lett. {\bf 87}, 227202 (2001).

\bibitem{model_dietl2}
T. Dietl, H. Ohno and F. Matsukura, Phys. Rev. B  {\bf 63}, 195205 (2001).

\bibitem{cdgep2}
P. Mahadevan and A. Zunger, Phys. Rev. Lett. {\bf 88}, 047205 (2002).

\bibitem{baldereschi}
A. Baldereschi and N.O. Lipari, Phys. Rev. B {\bf 8}, 2697 (1973).

\bibitem{old_dietl}
A. Haury, A. Wasiela, A. Arnoult, J. Cibert, S. Tatarenko, T. Dietl, Y. Merle d. Aubigne, 
Phys. Rev. Lett. {\bf 79}, 511 (1997).

\bibitem{ferro_insulating} 
F. Matsukura, H. Ohno, A. Shen and Y. Sugawara, Phys. Rev. B {\bf 57}, R2037 (1998);
H. Ohno and F. Matsukura, Solid State Commun. {\bf 117}, 179 (2001).

\bibitem{kondoiivi}
B.E. Larson, K. C. Hass, and H. Ehrenreich, A. E. Carlsson, Phys. Rev. B {\bf 37}, 4137 (1988).

\bibitem{ihmzunger}
J.~Ihm, A.~Zunger and M.L.~Cohen, J. Phys. C:{\bf 12}, 4409 (1979).

\bibitem{usp}
D. Vanderbilt, Phys. Rev. B {\bf 41}, 7892 (1990);

\bibitem{vasp}
G.~Kresse and J.~Furthm$\ddot{u}$ller, Phys. Rev. B. {\bf 54}, 11169 (1996);
G.~Kresse and J.~Furthm$\ddot{u}$ller, Comput. Mat. Sci. {\bf 6}, 15 (1996).

\bibitem{fiorentini}
V. Fiorentini, M. Methfessel and M. Scheffler, Phys. Rev. B {\bf 47}, 13353 (1993).

\bibitem{pw91}
J.P. Perdew and W. Wang, Phys. Rev. B {\bf 45}, 13244 (1992).

\bibitem{expt_lattice}
GaN: T.Lei, M. Fanciulli, R.J. Molnar, T.D. Moustakas, R.J. Graham, J. Scanlon, Appl. Phys. Lett. {\bf 59}, 944 (1991);
M.J. Paisley, Z. Sitar, J.B. Posthill, R.F. Davis, J. Vac. Sci. Technol. A {\bf 7}, 701 (1989);
GaP: V.N. Bessolov, T.T. Dedegkaev, A.N. Elfimov, N.F. Kartenko, Yakovlev, P. Yu, Sov. Phys. Solid State
{\bf 22}, 1652 (1980); GaAs: J.B. Mullin, B.W. Straughan, C.M.H. Driscoll, A.F.W. Willoughby, Inst. Phys. Conf. Ser. {\bf 24}, 275 (1975);
GaSb: M.E. Straumanis and C.D. Kim, J. Appl. Phys. {\bf 36}, 3822 (1965).

\bibitem{capz}
J. Perdew and A. Zunger, Phys. Rev B {\bf 23}, 5075 (1981).

\bibitem{cuinse2}
S.B.~Zhang, S.H.~Wei and A.~Zunger, Phys. Rev. Lett. {\bf 78}, 4059 (1997);
S.B.~Zhang, S.H.~Wei, A.~Zunger and H. Katayama-Yoshida, Phys. Rev. B. {\bf 57}, 9642 (1998); 

\bibitem{payne}
G. Makov and M.C. Payne, Phys. Rev. B {\bf 51}, 4014 (1995).

\bibitem{landolt}
{\it Semiconductors Basic Data}, edited by O. Madelung, Springer-Verlag (1996).

\bibitem{mnte}
S.H.~Wei and A.~Zunger, Phys. Rev. Lett. {\bf 56}, 2391 (1986); {\it ibid.}, Phys. Rev. B {\bf 37}, 8958 (1988).

\bibitem{cef2}
O. Eriksson, L. Nordstrom, M.S.S. Brooks and B. Johansson, Phys. Rev. Lett. {\bf 60}, 2532 (1988).

\bibitem{sfmo}
D.D.~Sarma, P. Mahadevan, T. Saha-Dasgupta, S. Ray and A. Kumar, Phys. Rev. Lett. {\bf 85}, 2549 (2000);
D.D. Sarma, Curr. Opinion in Solid State and Mat. Sc {\bf 5}, 261 (2001).

\bibitem{boffset}
S.H.~Wei and A.~Zunger, Appl. Phys. Lett. {\bf 72}, 2011 (1998).

\bibitem{fazzio}
M.J. Caldas, A. Fazzio and A. Zunger, Appl. Phys. Lett. {\bf 45}, 671 (1984).

\bibitem{dietl_rev} 
T. Dietl, Semicon. Sci. and Tech. {\bf 17}, 377 (2002).

\bibitem{fujimori}
J.~Okabayashi {\it et al.}, Phys. Rev. B {\bf 58}, 4211 (1999).

\bibitem{sicref}
A. Zunger, Phys. Rev. Lett. {\bf 50}, 1215 (1983).

\bibitem{zungeriivi}
S.B. Zhang, S.H. Wei, and A. Zunger, Phys. Rev. B {\bf 52}, 13975 (1995).

\bibitem{optical_prl}
E.J. Singley {\it et al.}, Phys. Rev. Lett. {\bf 89}, 097203 (2002).

\bibitem{baraff}
G.B.~Bachelet, G.A.~Baraff and M.~Schluter, Phys. Rev. B {\bf 24}, 915 (1981).

\bibitem{balhausen}
C.J.~Balhausen, {\it Ligand Field Theory}, McGraw-Hill, New York (1962).

\end{thebibliography}

\newpage

\begin{table}
\caption
{ Comparison of GGA optimised lattice constants with experiment for the pure host}
\begin{tabular}{c|c|c}
System & experiment a (in $\AA$) & GGA PW 91 a(in $\AA$) \\ \hline
GaN    & a=4.49; 4.51 & a=4.53 \\ \hline
GaP    & a=5.45             & a=5.489           \\ \hline
GaAs   & a=5.65             & a=5.728             \\ \hline
GaSb   & a=6.10             & a=6.18              \\ \hline  
\end{tabular}
\end{table}

\begin{table}
\caption
{ Impurity formation energies with and without Makov-Payne charge corrections for the acceptor transitions for
3$d$ impurities in GaAs. Experimental transitions have been given in brackets for comparison.
}
\begin{tabular}{l|l|l}
System & Formation energies & Formation energies \\
       & (no charge correction) & (with charge correction)  \\ \hline
GaAs:V (q=0) & 1.22 +$\mu_{Ga}$ - $\mu_V$                & 1.22 \\
GaAs:V (q=+1)& 1.32 +$\mu_{Ga}$ - $\mu_V$ + $\epsilon_F$ & 1.41 \\
GaAs:V (q=-1)& 2.05 +$\mu_{Ga}$ - $\mu_V$ - $\epsilon_F$ & 2.15 \\
GaAs:V (q=-2)& 3.18 +$\mu_{Ga}$ - $\mu_V$ - 2$\epsilon_F$ & 3.56 \\
             & (0/-) = 0.83 eV                            & (0/-) = 0.93 eV         \\ 
             & (-/2-) = 1.13 eV                           & (-/2-) = 1.41 eV              \\ \hline
GaAs:Cr (q=0) & 1.61 +$\mu_{Ga}$ - $\mu_{Cr}$             & 1.61 \\
GaAs:Cr (q=+1)& 1.47 +$\mu_{Ga}$ - $\mu_{Cr}$ +$\epsilon_F$  & 1.56 \\
GaAs:Cr (q=-1)& 2.02 +$\mu_{Ga}$ - $\mu_{Cr}$ -$\epsilon_F$  & 2.115\\
GaAs:Cr (q=-2)& 2.81 +$\mu_{Ga}$ - $\mu_{Cr}$ -2$\epsilon_F$  & 3.20 \\
             & (0/-) = 0.41 (0.74) eV                            & (0/-) = 0.51 (0.74) eV              \\ 
             & (-/2-) = 0.79 (1.57) eV                           & (-/2-) =1.09 (1.57) eV              \\ \hline
GaAs:Mn (q=0) & 1.04 +$\mu_{Ga}$ - $\mu_{Mn}$             & 1.04 \\
GaAs:Mn (q=+1)& 1.15 +$\mu_{Ga}$ - $\mu_{Mn}$ +$\epsilon_F$  & 1.24 \\
GaAs:Mn (q=-1)& 1.13 +$\mu_{Ga}$ - $\mu_{Mn}$ -$\epsilon_F$  & 1.23 \\
             & (0/-) = 0.09 (0.11) eV                            & (0/-) = 0.19 (0.11) eV              \\  \hline
GaAs:Fe (q=0) & 1.79 +$\mu_{Ga}$ - $\mu_{Fe}$             & 1.79 \\
GaAs:Fe (q=+1)& 1.83 +$\mu_{Ga}$ - $\mu_{Fe}$ +$\epsilon_F$  & 1.92 \\
GaAs:Fe (q=-1)& 2.10 +$\mu_{Ga}$ - $\mu_{Fe}$ -$\epsilon_F$  & 2.21 \\
GaAs:Fe (q=-2)& 2.82 +$\mu_{Ga}$ - $\mu_{Fe}$ -2$\epsilon_F$  & 3.27 \\
             & (0/-) = 0.31 eV                            & (0/-) = 0.42 eV              \\ 
             & (-/2-) = 0.72 eV                           & (-/2-) = 1.06 eV              \\ \hline
GaAs:Co (q=0) & 1.84 +$\mu_{Ga}$ - $\mu_{Co}$             & 1.84 \\
GaAs:Co (q=+1)& 1.90 +$\mu_{Ga}$ - $\mu_{Co}$ +$\epsilon_F$  & 1.99 \\
GaAs:Co (q=-1)& 1.92 +$\mu_{Ga}$ - $\mu_{Co}$ -$\epsilon_F$  & 2.01 \\
             & (0/-) = 0.08 (0.16) eV                            & (0/-) = 0.17 (0.16) eV    \\  \hline
\end{tabular}
\end{table}

\begin{table}
\begin{tabular}{l|l|l}
System & Formation energies & Formation energies \\
       & (no charge correction) & (with charge correction)  \\ \hline
GaAs:Ni (q=0) & 1.73 +$\mu_{Ga}$ - $\mu_{Ni}$             & 1.73 \\
GaAs:Ni (q=+1)& 1.76 +$\mu_{Ga}$ - $\mu_{Ni}$ +$\epsilon_F$  & 1.86 \\
GaAs:Ni (q=-1)& 1.86 +$\mu_{Ga}$ - $\mu_{Ni}$ -$\epsilon_F$  & 1.96 \\
GaAs:Ni (q=-2)& 2.27 +$\mu_{Ga}$ - $\mu_{Ni}$ -2$\epsilon_F$  & 2.68 \\
             & (0/-) = 0.13 (0.22) eV                            & (0/-) = 0.23 (0.22) eV        \\ 
             & (-/2-) = 0.41 (1.13) eV                           & (-/2-) = 0.72 (1.13) eV        \\ \hline
\end{tabular}
\end{table}

\begin{table}
\caption
{The calculated energy-minimizing 
configuration for neutral substitutional
3$d$ impurities in GaAs. CFR states are given in square brackets and DBH states in round brackets. Boldface letters denote the 
first unoccupied orbital. Also shown are ground state multiplet  and in parentheses, local moment
$\mu_{loc}$ within a sphere of radius 1.2$\AA$ for isolated impurities. The last 2 columns give 
the total energy difference ${\Delta}E$ between 
FM and AFM spin arrangements
of TM pairs at first (NN) and fourth neighbor (4NN).
Asterisk denotes  the configuration with lowest energy. 
}
\begin{tabular}{c|c|c|c|c}
TM&{Configuration}&{Multiplet}&{$\Delta{E}_{NN}$}&{$\Delta{E}_{4NN}$}\\
~ &{}&($\mu_{loc}$)&(in meV) &(in meV) \\ \hline
Ni&  &  (0.53)&+2.85 &+4.3$^*$ \\ \hline
Co&  $[t_+^3 e_+^{2} t_-^0 e_-^{2}]$ $(t_-^3{\bf t_+^2})$ &$^3T_2$~(1.58)& -9.6$^*$ &-22.6\\ \hline
Fe&  $[t_+^3 e_+^{2} t_-^0 {\bf e_-^{0}}]$ $(t_-^3t_+^3)$ &$^6A_1$~(3.27)& +298$^*$ &+205\\ \hline
Mn&  $[t_+^3 e_+^{2} t_-^0 e_-^{0}]$ $(t_-^3{\bf t_+^2})$ &$^5T_2$~(3.75)& -247$^*$ &-227\\ \hline
Cr&  $[t_+^3 e_+^{2} t_-^0 e_-^{0}]$ $(t_-^3{\bf t_+^1})$ &$^4T_1$~(2.99)& -315$^*$ &-258\\ \hline
V &  $[{\bf t_+^0} e_+^{2} t_-^0 e_-^{0}]$ $(t_-^3t_+^3)$ &$^3A_2$~(1.84)& -40 &+31$^*$\\ \hline
V$^-$ &  $[{\bf t_+^0.5} e_+^{2} t_-^0 e_-^{0}]$ $(t_-^3t_+^3)$ & & -204 &\\ \hline
Fe$-$&  $[t_+^3 e_+^{2} t_-^0 {\bf e_-^{0.5}}]$ $(t_-^3t_+^3)$ & & +259$^*$ & \\ \hline
\end{tabular}
\end{table}

\begin{table}
\caption { Spin-up and spin down eigenvalues at $\Gamma$ for 3$d$ 
impurities in GaN, GaP and GaAs, referenced to the valence band 
maximum of the pure host. $\Delta$ corresponds to the spin splitting between up and down spin states.}
\begin{tabular}{l|l|l|l}
System & $E_v^+$ (eV)& $E_v^-$  (eV)& $\Delta$ (eV)\\ \hline
GaAs:V & -0.14 & -0.08 & 0.06 \\
GaAs:Cr & -0.26 & -0.08 & 0.18 \\
GaAs:Mn & -0.47 & -0.08 & 0.39 \\
GaP:Cr  & -0.22 & -0.08 & 0.14 \\
GaP:Mn  & -0.39 & -0.09 & 0.30 \\
GaN:Cr  & -0.07 & -0.03 & 0.04 \\
GaN:Mn  & -0.13 & -0.03 & 0.10 \\ \hline
\end{tabular}
\end{table}

\end{document}